\pdfoutput=1
\documentclass[
  twocolumn,
  prd,
  showpacs,
  amsmath,
  amssymb,
  superscriptaddress,floatfix
]{revtex4-1}
\usepackage{amsfonts,graphics,graphicx,epsfig,color,times,indentfirst,layout,hyperref,subfigure,multirow,bm,ulem}
\usepackage{braket}
\usepackage{pifont}
\usepackage{empheq}
\usepackage{scalerel,stackengine}
\hypersetup{linktocpage,,colorlinks=true,citecolor=cyan}
\stackMath
\newcommand\reallywidehat[1]{%
\savestack{\tmpbox}{\stretchto{%
  \scaleto{%
    \scalerel*[\widthof{\ensuremath{#1}}]{\kern-.6pt\bigwedge\kern-.6pt}%
    {\rule[-\textheight/2]{1ex}{\textheight}}
  }{\textheight}%
}{0.5ex}}%
\stackon[1pt]{#1}{\tmpbox}%
}
\newcommand{\tr}{\text{tr}}
\renewcommand{\v}[1]{\textbf{\textit #1}}
\newcommand{\vk}{\textbf{\textit k}}

\newcommand{\dg}{^{\dagger}}

\newcommand{\pmat}[1]{\begin{pmatrix}#1\end{pmatrix}}

\newcommand{\bk}{{\bf k}}

\newcommand{\vx}{{\v x}}
\newcommand{\vy}{{\v y}}

\begin{document}

\title{The Coulomb problem in iron based superconductors}

\author{Elio J.\ K\"onig}
\affiliation{Department of Physics and Astronomy, Rutgers University, Piscataway, New Jersey, 08854, USA}

\author{Piers Coleman}
\affiliation{Department of Physics and Astronomy, Rutgers University, Piscataway, New Jersey, 08854, USA}
\affiliation{Department of Physics, Royal Holloway, University of London, Egham, Surrey TW20 0EX, UK}
\date{\today}

\begin{abstract}
We discuss the role of strong Coulomb interactions 
in iron-based superconductors (FeSCs). 
The presumed s$^\pm$ character of these
superconductors means that the condensate is not symmetry protected
against Coulomb repulsion.  
Remarkably, the transition temperatures and the excitation gap
are quite robust across the large
family of iron based superconductors, 
despite drastic changes in Fermi surface geometry.
The Coulomb problem is to understand 
how these superconductors
avoid the strong onsite Coulomb interaction at the iron atoms, while
maintaining a robust transition temperature. 
Within  the dominant space of
$t_{2g}$ orbitals, on-site repulsion in the FeSCs 
enforces two linearly independent
components of the condensate to vanish. 
This raises the possibility
that iron-based superconductors 
might adapt their condensate to the Coulomb constraints
by rotating the pairing state within the large 
manifold of entangled, extended s-wave gap functions with different orbital and
momentum space structure. 
We examine this ``orbital and k-space flexibility'' (OKF) mechanism 
using both Landau theory and microscopic calculations
within a multi-orbital t-J model.  Based on our results, 
we conclude that OKF {necessitates}
 a large condensate degeneracy.  
One interesting possibility raised by our results, 
is that a resolution to the Coulomb problem in FeSC
might require a reconsideration of triplet pairing. 
\end{abstract}


\maketitle

\section{Introduction} Ten years after their
discovery~\cite{KamiharaWatanabe2008}, iron based superconductors
(FeSC) remain at the focus of condensed matter research. These
materials offer  great promise for applications, 
providing robust high temperature superconductivity at high current
densities and magnetic fields. On the other hand, from a fundamental physics
perspective, the rich phase diagram and the 
unresolved pairing mechanism of the iron-based superconductictors
continue to attract both experimental and theoretical attention.

Discoveries over the past decade have revealed a broad family 
of iron-based superconductors. 
Generally, they are
categorized by their chemical compositions, and the most prominent
classes are the 1111, 122, 111, and 11 families.   
All iron-based superconductors share a common local electronic
structure, with the iron atoms contained within a tetrahedron
of pnictide or chalgonide atoms. 
These tetrahedra are closely packed in a staggered formation with two
iron atoms per unit cell, forming 
a two dimensional layered structure.
There are two slight exceptions: 
the quasi-1D compound 
BaFe$_2$S$_3$~\cite{TakahashiOhgushi2015,YamauchiOhgushi2015} in which
the paired tetrahedra form two-leg ladders of iron atoms, and 
single layer FeSe~\cite{HuangHoffmann2017}. 
By contast, the momentum-space electronic structures of the
FeSC show great diversity, see Table ~\ref{tab:FSstructureandTc}. Many of
the layered materials display both hole and electron pockets,
respectively located
at the $\Gamma$ and $X$ points of the Brillouin zone.
However, KFe$_2$As$_2$ only has hole pockets centered at
the $\Gamma$ point~\cite{SatoDing2009,YoshidaHarima2011}, while in
various iron chalcogenides there are only electron pockets
~\cite{ChangWu2015}. There is no indication for a clear trend in the transition temperatures as pockets disappear. At the same time, a universal $2\Delta_{\rm max}/k_B T_c \approx 7.2$ suggests a common mechanism for superconductivity in all FeSC~\cite{LeeKotliar2018}.

The Fermi surfaces of the FeSC are primarily 
composed from the $t_{2g}$ orbitals of the iron
$d$-shell. The electron correlations increase from the 1111 to 122 and 111
compounds, and are maximal in the 11 materials. First
principle calculations~\cite{MiyakeImada2010} indicate that the 
intra orbital and inter
orbital on-site Coulomb energies, denoted by $U$ and $U'$, are about 5 to
15 times larger than the hopping matrix elements, which are comparable
to the Hund's coupling $J_H$. Evidence for Hubbard-like
side bands~\cite{WatsonValenti2017,EvtushinskyBuchner2016}, were reported
very recently, however, the vast majority of FeSC do not
display a Mott phase. The strong interaction energies on the one hand,
and the experimentally observed metallicity on the other hand have
catalyzed two opposing theoretical view points,  which focus on
``local''~\cite{deMedici2015, SiAbrahams2016} and
``itinerant''~\cite{Chubukov2015, FernandesChubukov2016} aspects of
the physics. 

Based on Knight shift experiments, the superconducting gap
structure~\cite{Hirschfeld2016} of iron based superconductors 
is believed to be spin-singlet in character. 
While many materials show a full gap, there is some indirect
experimental support for condensate pairing amplitudes with  opposite
signs on different Fermi surfaces~\cite{BangStewart2017}.  Recent
experiments have underlined the importance of orbital selective
physics, both in the normal state~\cite{HardyMeingast2013,YiLu2015}
and in the superconducting state~\cite{SprauSeamusDavis2017}.

\subsection{The Coulomb Problem}\label{}

Despite the enormous progress of the past years, 
two unsolved questions about the 
pairing mechanism in the iron-based superconductors stand out:
\begin{itemize}

\item Why does
superconductivity appear so generically, independently of very
different Fermi surface topologies? 

\item What is the common mechanism by which FeSC
overcomes the strong Coulomb repulsion at the iron sites?
\end{itemize}
In almost all other 
strongly correlated pair condensates
including the 
cuprate superconductors, Sr$_2$RuO$_4$, heavy fermion and organic
superconductors, and superfluid $^3$He, the condensate
avoids the local repulsion by forming finite angular momentum Cooper pairs.
This is because the nodes in the finite angular momentum
Cooper pairs guarantee a vanishing on-site
component of the pair expectation value, thereby protecting the
condensate against large Coulomb repulsion. 

\begin{widetext}
In an orbital basis, the onsite Coulomb interaction
is written
\begin{eqnarray}\label{l}
\hat V_{C} =
\frac{1}{2}\sum_{j, m_{i},\alpha,\beta } (m_{1},m_{2}|\hat
V|m_{1}',m_{2}')
\psi \dg_{m_{1}\alpha } (\v x_{j})
\psi \dg_{m_{2}\beta} (\v x_{j})
\psi _{m_{2}'\beta} (\v x_{j})
\psi _{m_{1}'\alpha} (\v x_{j}),
\end{eqnarray}
$\psi\dg _{ m, \alpha}(\v x_{j})$ 
creates an electron in orbital $m$, spin component
$\alpha\in \{\uparrow,\downarrow  \}$ at position $\v x_{j}$ and
\begin{equation}
(m_{1},m_{2}|\hat
V|m_{1}',m_{2}')  = \int d^{3}xd^{3}y V_{c} (\vx-\vy)
\phi_{m_{1}} (\vx)\phi_{m_{1}'} (\vx)
\phi_{m_{2}} (\vy) \phi_{m_{2}'} (\vy)
\end{equation}
is the integral of the Coulomb interaction $V_{c} (\vx-\vy)$ 
with the atomic orbitals, $\phi_{m}(\vx)$.  In an isotropic
environment, a simplified version of the interaction can be  written\cite{footnote}
\begin{equation}
\hat V =
\frac{U}{2}\sum_{j}n_{j}^{2}- J_{H,1} \sum_{j}\vec{ S}_{j}^{2}-J_{H,2}\sum_{j}\vec{L}_{j}^{2} \label{eq:HKanamori}
\end{equation}
where $n_{j}= \sum_{m,\alpha } \psi \dg_{m\alpha } (\vx_{j})\psi_{m\alpha }
(\vx_{j})$ is the number of d-electrons at site $j$, while $\vec{S}_{j}$ and
$\vec{L}_{j}$ are the corresponding total spin and orbital angular
momentum at site j, respectively. The leading Coulomb 
repulsive $U$ is of order 2-4eV, while the subleading 
Hund's interactions, $J_{H,1}$ and $J_{H,2}$ are typically a few tenths of eV~\cite{MiyakeImada2010}.
These large interactions are certainly not unique to the iron-based
superconductors: they feature in almost all iron-based compounds
and they drive a wide variety of both Mott and Hund's physics, such as
the ferromagnetism in iron, which develops at the astonishingly high
temperature of 1043K, and the Mott 
insulating behavior of rust (Fe$_{2}$O$_{3}$).
\end{widetext}

The important point is that in the iron-based superconductors, the
characteristic gap energies on the scale of tens of millivolts, 
are dwarfed by the onsite Coulomb interactions on the scale of volts.
Once a superconducting condensate forms, the additional
onsite charge fluctuations associated with the coherent state
modify the Coulomb energy.  The key quantity determining this condensate
correction to the Coulomb energy is the anamolous equal-time Gor'kov
function
\begin{equation*}\langle \psi_{ m \alpha}(\v x_{j})
\psi_{ m' \beta}(\v x_{j}) \rangle  = \epsilon_{\alpha \beta} {F}_{mm'}
 \label{eq:Gor'kov}.
\end{equation*}
Here we 
have restricted ourselves to spin-singlet pairing, in which the
Gor'kov function is proportional 
to the antisymmetric tensor $\epsilon_{\alpha \beta
}=-\epsilon_{\beta \alpha }$. If we evaluate the change in the
condensation energy in the Hartree Fock approximation, by contracting
the anomalous terms in the energy, then for the isotropic interaction
the change in condensate energy per site is
\begin{align}
\Delta E &= \frac{1}{N_{s}}\left[ 
\langle \psi_{SC}|\hat V\vert \psi_{SC}\rangle - 
\langle \psi_{FL}|\hat V\vert \psi_{FL}\rangle 
\right]\cr
&
=\sum_{m,m'}|F_{m,m'}|^{2} [\tilde U  + 2 J_{H,2}\delta_{m,m'}] ,
\end{align}
where $N_{s}$ is the number of sites, and $\ket{\Psi_{SC}}$ ($\ket{\Psi_{FL}}$) is the superconducting (Fermi liquid) many body ground state.
Here, the Coulomb cost is $\tilde U = U+
({3}J_{H,1}-4J_{H,2})/{2}$, to be specific 
we have considered the case of a 
$t_{2g}$ triplet ($m\in \{zx,zy,xy\}$).
The huge discrepancy of scale between the Coulomb and gap energies, means that
for the stabilization of the condensate, the onsite pairing terms have to
vanish, i.e 
\begin{equation}
F_{m,m'}= \frac{1}{2}\epsilon_{\beta\alpha } \left \langle \psi_{ m \alpha}(\v x_{j})
\psi_{ m' \beta}(\v x_{j})\right \rangle  = 0.
 \label{eq:VanishingPairCorrelator}
\end{equation}
These are the Coulomb constraints on the condensate. 

If we rewrite the Coulomb constraint 
in the momentum and frequency domain it
becomes
\begin{equation}\label{eq:integral}
F_{m,m'} = \int \frac{d^{3}k}{(2\pi)^{3}}\frac{d\omega}{\pi} f (\omega) {\rm Im}[F_{m,m'}
(\vk ,\omega-i\delta )] = 0, 
\end{equation}
where $f (\omega)= (\exp[\beta \omega]+1 )^{-1}$ is the Fermi function.
Typically, the integrand in 
Eq.~\eqref{eq:integral} 
is dominated by the Fermi
surfaces and is quite sensitive to the electronic structure.
In d-wave superconductors, such as the heavy fermion and cuprate
superconductors, satisfaction of the Coulomb constraint is 
symmetry-protected, 
because the alternating signs of the four
quadrants of the gap function cause the momentum summation to
automatically vanish.  
However, for s-wave superconductors the Coulomb constraint is unprotected by symmetry. 
In many of the early iron-based
superconductors the Coulomb constraint was thought to be satisfied by
dint of the $s_{\pm}$ character, with alternating gap signs on the
electron and hole pockets.  However, the subsequent discovery that
iron-based superconductivity persists 
in a wide class of materials without electron, or without hole
pockets raises the question of how the Coulomb constraint is satisfied
without loss of superconducting transition temprature.  

In fact crystal symmetries do protect the {\sl off-diagonal} Coulomb
constraints. Within the subspace of  
dominant $t_{2g}$ orbitals,  
there are \`a priori six independent
constraints.  However, 
the off-diagonal  components  of $F$
change sign under some of the non-symmorphic crystal symmetry
operations and since an s-wave condensate is invariant
under the crystal symmetries, 
the off-diagonal components of $F$ must
vanish,  $F_{mm'}=0$ ($m\neq m'$).  For example, 
under a combined $\pi/2$ rotation about the $z-$
axis and mirror-reflection in
the xy plane,  $(x,y,z)\rightarrow (y,-x,-z)$ and so 
$F_{zx,zy}\rightarrow -F_{zy,zx}$, which
must then vanish for an s-wave condensate. 
Tetragonal crystalline symmetry
also guarantees that $F_{zx,zx}= F_{zy,zy}$, are equal, so 
the Coulomb repulsion thus imposes two independent
local constraints:  
\begin{eqnarray}\label{localcons}
F_{zx,zx}&=&F_{zy,zy}= 0,\cr
F_{xy,xy}&=&0.
\end{eqnarray}

In conventional phonon-paired superconductors the 
bare Coulomb repulsion is screened by
virtual high energy pair
fluctuations~\cite{McMillan1968,BogolyubovShirkov1959}.  This pair-screening
process is logarithmically slow in energy, but the
exponential separation of time-scales between
the instantaneous Coulomb interaction and highly retarded electron-phonon
interaction is sufficient to allow the electron-phonon interaction to
win at low energies.  This pair screening effect actually imposes a
sign-change between the high- and low- frequency components of the
Gor'kov function to satisfy the onsite equal-time constraint. 
In a purely-electronic pairing mechanism relevant
to the FeSC, such a clean separation of time-scales is absent, so we can
not appeal to retardation to impose the Coulomb constraint.

As such, the Coulomb problem is conceptually equal for FeSC with tetragonal or orthorhombic symmetry. We here concentrate on the case of fourfold rotational symmetry and mention that the orthorhombic symmetries also ensure vanishing off-diagonal components $F_{mm'} \vert_{m \neq m'}$ leading to three independent Coulomb conditions $F_{mm} = 0$. 

\begin{table}
\begin{tabular}{|p{.155\textwidth}|p{.07\textwidth}|p{.053\textwidth}|p{.104\textwidth}|p{.06\textwidth}|}
\hline
Compound & FS & $T_c$ & & Ref. \\
\hline \hline
monolayer FeSe & 0h, 2e & 65 K$^\sharp$ & on SrTiO$_3$ & \cite{HeLi2013}\\
\hline
Ba$_{1-x}$K$_x$Fe$_2$As$_2$ & 2h, 2e& 37 K & @ x = 0.4 &\cite{EvtushinskyBorsenko2009,DingFedorov2011} \\
\hline
K$_{x}$Fe$_{2-y}$Se$_2$ & 0h, 2e & 30 K & @ (x,y) = (0.8, 0.3) & \cite{QianDing2011}\\
\hline
BaFe$_2$S$_3$ & quasi 1D & 24 K & @ 10.6 GPa & \cite{TakahashiOhgushi2015, YamauchiOhgushi2015}\\
\hline
Ba(Fe$_{1-x}$Co$_x$)$_{2}$As$_2$ & 2h$^\dagger$, 2e & 22 K & @ x = 0.058 & \cite{LiuKaminski2010,LiuKaminski2011}\\
\hline
Fe$_{1+y}$Se$_x$Te$_{1-x}$& 3h, 2e & 14.5 K & @ x = 0.45  & \cite{TamaiBaumberger2010, LiuShen2015}\\
\hline
LaOFeP & 2h, 2e & 6K & & \cite{ColdeaMcDonald2008}\\
\hline
KFe$_2$As$_2$ & 3h$^*$, 0e & 4K &  &\cite{YoshidaHarima2011}\\
\hline
\end{tabular} 
\caption{Summary of Fermisurface (FS) structure and $T_c$ for a multitude of FeSC materials. The notation ``$m$h, $n$e'' means $m$ hole pockets centered around the $\Gamma$ point and $n$ electron pockets. Footnotes: $^*$ There are additional 4 small hole pockets which are not centered around a high symmetry point. $^\dagger$ 3D (i.e. non-cylindrical) character of the Fermi sheets, is important at the $\Gamma$ point. $^\sharp$ $T_c$ up to 109 K was reported in this material~\cite{HuangHoffmann2017}. All references are based on ARPES, except ~\cite{ColdeaMcDonald2008}, which is quantum oscillations, and ~\cite{TakahashiOhgushi2015, YamauchiOhgushi2015}. We are not aware of an experimental study of the electronic structure of BaFe$_2$S$_3$ (this material is insulating at ambient pressure). The ARPES data of these materials indicates a largest Fermi energy of $40$ to $150$ meV.}
\label{tab:FSstructureandTc}
\end{table}

\subsection{Multiband Kohn-Luttinger physics: RG} \label{sec:RG} A
widely proposed mechanism for superconductivity from repulsive
interactions relies on a generalization of the Kohn-Luttinger idea to
multiband systems, often explored using renormalization group (RG)
approaches~\cite{MaitiChubukov2010,ChubukovKhodasFernandes2016,ThomaleBernevig2009,KoenigColeman2017}. These
approaches have been used to argue that under a wide range of multi-band
circumstances, the renormalization group flow develops an attractive
pairing instability that overcomes the Coulomb constraint. 

{To illustrate these arguments and their connection to the Coulomb
problem, we focus on two-band models. In this situation, the ladder
resummation of logarithmic corrections to interband (intraband) Cooper
coupling constants $G_{12}$ ($G_{11}, G_{22}$) is expressed in terms of
two RG equations (see also App.~\ref{app:RG})
\begin{equation}
\frac{d g_\pm}{d \ln(D/T)} = - g_\pm^2, \label{eq:RGeqs}
\end{equation} 
where we introduced the cutoff $D$, temperature $T$ and 
\begin{equation}
 g_\pm = \frac{G_{11} \rho_1 + G_{22} \rho_2 \pm \sqrt{4 G_{12}^2 \rho_1 \rho_2 + (G_{11} \rho_1 - G_{22} \rho_2)^2}}{2}.
\end{equation}
Here $\rho_{1,2}$ denote density of states of the respective bands.
While $g_+ > 0$ generically for repulsive interactions, $g_-$ 
becomes negative 
for sufficiently strong interband interaction and its runaway
flow signals the onset of $s_\pm$ pairing. Importantly, $g_- < 0$ can
be induced dynamically by fluctations in the particle-hole channel
(Kohn-Luttinger mechanism).  Since the presence of a Cooper
instability in Eq.~\eqref{eq:RGeqs} holds even for strong repulsion,
these arguments suggest that the weak-coupling 
RG provides a resolution to the Coulomb problem. }

However, a closer inspection reveals a difficulty with this line
of reasoning. 
Each momentum space RG theory is tailor-made
for a specific subset of FeSC compounds with the same Fermi surface
topology, {as such it} does not provide a generic explanation for the 
robustness of the transition temperature  against the appearance and
disappearance of electron and hole pockets. 
Even within a family with a fixed Fermi surface
topology the critical temperature suffers from unrealistically large
variations as the relative carrier concentration in different pockets
changes from compound to compound. We can illustrate this with the
the two-band model. Suppose we concentrate on the limit of infinite
repulsions where $G_{11} = G_{22} = G_{12} - g = U \rightarrow
\infty$. While the robust existence of a Cooper instability in RG
demonstrates an asymptotic orthogonalization against Coulomb
repulsion, the microscopic on-site constraint, Eq.~\eqref{localcons},
is beyond its reach. In practice, this is reflected by a strong dependence of
the UV value of the coupling constant
\begin{equation}\label{eq:gminusbare}
g_-(0) = g\frac{\rho_1 + \rho_2}{2} (1-\delta^{2})
\end{equation}
on the relative density of states $\delta = (\rho_1 -\rho_2)/(\rho_1 +
\rho_2)$, which leads to exponential suppression of $T_c = D \exp[-1/g_-(0)]$ when $\delta$ is of order unity.
This strong suppression should be contrasted to the case of phononic BCS theory where, due to retardation, effects~\cite{McMillan1968,BogolyubovShirkov1959}, the effective Coulomb suppression $\mu^*$ is (i) weak and (ii) slowly dependent on UV physics.
We
conclude that RG only partially solves 
the Coulomb problem, and that 
search for a generic mechanism for Coulomb-protection is still unfulfilled. 

\begin{figure}
\includegraphics[scale=.25]{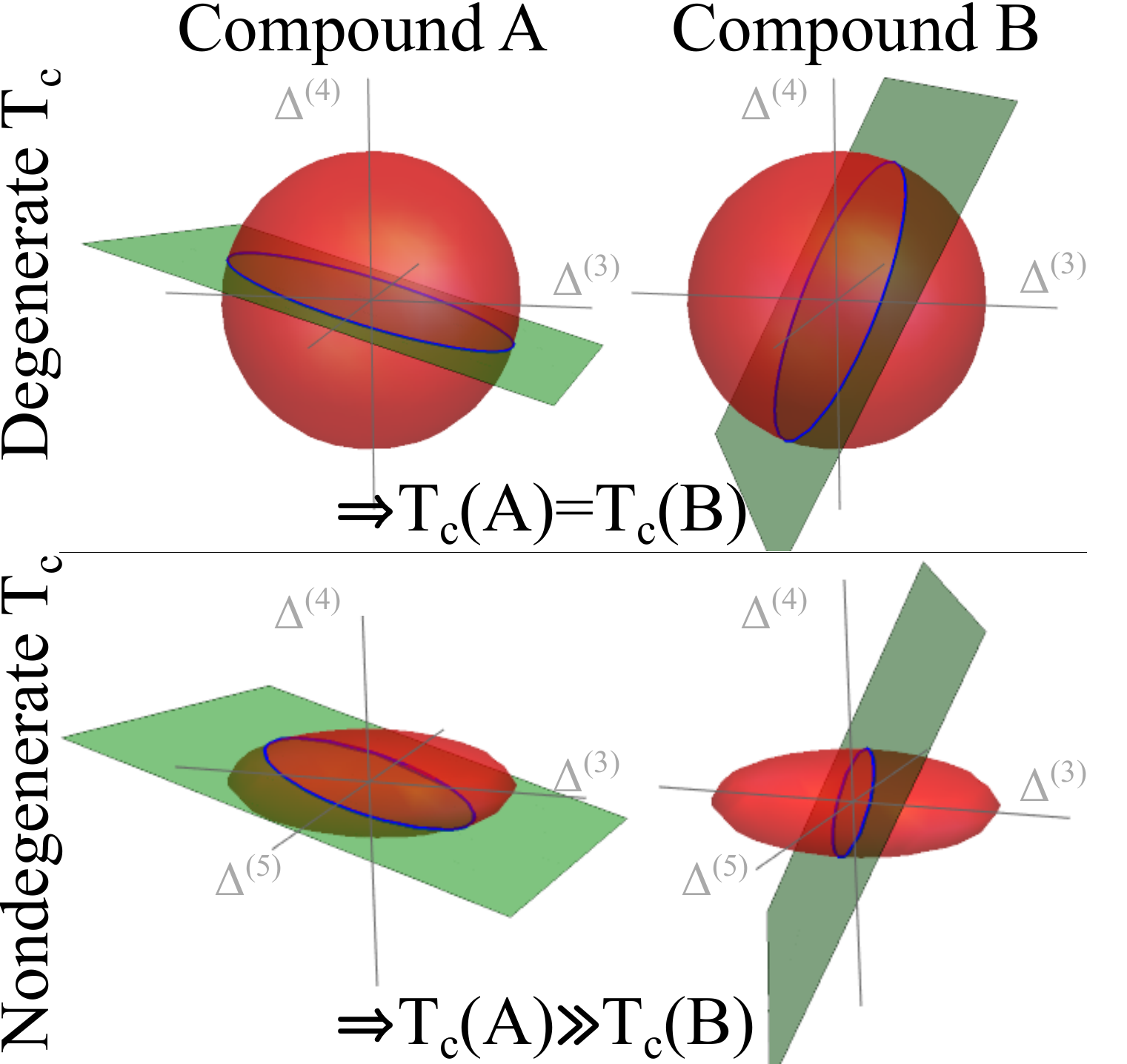}
\caption{Schematic illustrating orbital and k-space flexibility. Here, the large space of superconducting order parameters is represented as a three dimensional space. Each direction has an associated ``bare'' $T_c (\vec \Delta)$ which is determined in the fictitious case without on-site repulsion and which is represented as the distance from the origin. It varies continuously and thus forms a surface, e.g. a sphere or an ellipsoid. The orientation with maximal $T_c$ determines the pairing state. On-site Coulomb repulsion constraints the order parameter space to a submanifold, see Eq.~\eqref{eq:VanishingPairCorrelator}. The latter is material dependent and represented by a plane. Upper panels: When the bare $T_c$ is the same for all pairing directions, the orientation adapts to Coulomb repulsion without any cost in $T_c$ ("orbital and k-space flexibility"). Lower panels (realistic situation): The bare $T_c$ is direction dependent, orbital and k-space flexibility fails to explain comparable $T_c$ for different materials.}
\label{fig:Flexibility}
\end{figure}

\subsection{Orbital and k-space  Flexibility}

A central element of the iron based superconductors, 
is the orbital degrees of freedom. In general, the gap function
$\Delta_{m,m'} (\vk )\equiv \langle m \vk \vert \hat \Delta \vert m'\vk \rangle  $
is a momentum-dependent operator in the space of orbital quantum
numbers.  The simplest possibility is that pairing is orbitally trivial
$\Delta_{m,m'} (\vk )=\delta_{m,m'}\Delta (\vk
)$.  Another possibility, 
is orbitally selective pairing
~\cite{YinHauleKotliar2014,ChubukovFernandes2016,NicaSi2017}, in which
the gap function is orbitally diagonal, but changes sign 
dependent on the orbital content 
$\Delta_{m,m'} (\vk )=\delta_{m,m'}\Delta_{m} (\vk )$.  
The more general possibility, is {\sl orbital
entanglement}~\cite{GaoZhu2010,OngColeman2013,OngSchmalian2016}, in which
the gap function is an 
off-diagonal  matrix in the space of $d$-shell
orbitals. For example, 
\begin{equation}
\hat  \Delta \propto \ket{d_{xz}}\bra{d_{yz}}+\ket{d_{yz}}\bra{d_{xz}}.
\end{equation}   
In this case, the condensate wave function 
can not be written as a product state in the
space of orbitals, and is thus {\sl orbitally entangled}. 
Sizeable interorbital, spin singlet pairing was
predicted in material specific calculations on LiFeAs
~\cite{NourafkanTrembley2016} and LaOFeAs~\cite{WangMaier2015}.
Orbitally entangled spin triplet pairing has also been proposed for
LaFeAsO$_{1-x}$F$_{x}$~\cite{DaiZhang2008} and recently for
LaNiGa$_2$~\cite{WengYuan2016}.

The robust appearance of superconductivity in a wide variety of Fermi
surface morphologies, despite the absence of nodes in the gap to symmetry-protect against the Coulomb constraint, leads us to consider
the possibility that the condensate takes advantage of the large
number of s-wave pairing channels, 
adapting the orbital entanglement to minimize the Coulomb
interaction. 
Such {\sl orbital and k-space flexibility} (OKP)  would hypothetically 
allow the condensate to ``rotate'' within a manifold of almost
degenerate orbital pairing channels to satisfy the Coulomb
constraint, approximately preserving the transition temperature,
$T_{c}$ (see Fig.~\ref{fig:Flexibility}). 
In this scenario, as
the electronic structure changes from material to material, the
pairing channel flexibly rearranges in response to the 
the large space of Fermi-surface
morphologies. 

Motivated by these promising arguments, which are implicitly assumed in many theories of FeSC, 
we present a comprehensive study of
orbital and k-space flexibility as a way to solve the Coulomb
problem. 
Our principle conclusion, based on both phenomenological arguments and
microscopic calculations, is that OKF 
requires an internal degeneracy amongst the pairing states. The
apparent absence of such a degeneracy leads to a failure of orbital
flexibility, forcing us to reconsider the singlet pairing assumption, 
as we discuss in the conclusions. 

The paper is structured as follows:
Section~\ref{sec:PhenomenologicalLandau} is primarily
phenomenological, {and conceptually} explain{s} the Coulomb
constraints and the idea of 
orbital and k-space flexibility. 
Sec.~\ref{sec:Microscopics} provides {a systematic classification} of the superconducting matrix gap functions
within the space of $t_{2g}$ states {and} supporting, microscopic
calculations. We summarize
the derivation of the Landau theory and analyze a t-J model for a
family of electronic structures. We conclude with {a criticism and} an outlook, 
Sec.~\ref{sec:SummaryOutlook}, in which we discuss resolutions of the
Coulomb conundrum beyond the spin-singlet s-wave channel.

\section{Landau Theory}
\label{sec:PhenomenologicalLandau}

\begin{table}
\begin{tabular}{llcr}
$\Gamma$
&\;   Name & 
$\varphi^{\Gamma}(\v k) \lambda_{i(\Gamma)}$ &{irrep}    \notag \\
\hline
\hline
{1.} &\; { conventional s-wave}  & $\lambda_0$ & \;A$_{1g}$ \\ 
{2.} &\; { orbital-antiphase}  & $\lambda_8$ & \;A$_{1g}$\label{eq:OrbitalAntiphase} \\
\hline
{3.} &\; { conventional s$_\pm$ \cite{MazinDu2008}} & $c_x
c_y\lambda_0$ &\; A$_{1g}$
\\
{4.} &\; { orbitally-entangled \cite{GaoZhu2010,OngColeman2013,OngSchmalian2016}} &  $s_x s_y \lambda_1$& \;B$_{2g}$ \\
{5.} &\; { orbitally-selective}&  $(c_x-c_y)\lambda_3$ & \;B$_{1g}$\\
{6.} &\; { orbitally-antiphase
s$_\pm$\cite{YinHauleKotliar2014}}  & $c_xc_y \lambda_8$ &\; A$_{1g}$\\
\hline\notag
\end{tabular}
\caption{Summary of most relevant extended s-wave, spin singlet pairing channels as they appear in the expansion of the matrix gap function in irreducible representations of the point group $
\Delta_{m m'} (\vk)  = \sum_{\Gamma = 1}^{\mathcal D}
\Delta_\Gamma \varphi^{\Gamma}(\v k) [\lambda_{i(\Gamma)}]_{m m'}$ (Gell-Mann matrices are denoted $\lambda_i$). We employed the notation $c_x = \cos(k_x), \; s_x = \sin(k_x)$ etc.~for the harmonics in the unfolded tetragonal Brillouin zone.}
\label{tab:Pairingstates}
\end{table}

In this section we cast the Coulomb problem 
in the iron based superconductors as a phenomenological Landau
theory. 

\subsection{Coulomb repulsion in the free energy}

{Leaving microscopic details, a derivation and a thorough symmetry
analysis to the subsequent Sec.~\ref{sec:Microscopics}, we anticipate
that a real order parameter, viz.~a superconducting gap
$\Delta_\Gamma$, can be associated to each of the many pairing
channels within the subspace of spin-singlet, extended s-wave states,
see also Table ~\ref{tab:Pairingstates}. Furthermore, in the present study of a three orbital model in the tetragonal phase, symmetries impose that exactly two such order parameters are penalized by the Coulomb energy. They represent intraorbital on-site pairing, see also Eq.~\eqref{localcons}. }
If we expand the {Landau f}ree energy in powers of the
 order parameters {we obtain}
\begin{widetext}
\begin{equation}
\mathcal F  =- \left (
\Delta_{C}, 
\Delta_{P}
 \right )
\left (\begin{array}{cc}
{\frac{\delta_{CC'}}{U_C}} +\chi_{CC'} & \chi_{CP'} \\ 
 \chi_{PC'} & -\frac{\delta_{PP'}}{g_P} + \chi_{PP'}
\end{array} \right )
\left (\begin{array}{c}
\Delta_{C'} \\ 
\Delta_{P'}
\end{array} \right )
+ \beta_{\Gamma_1 \Gamma_2 \Gamma_3 \Gamma_4}  \Delta_{\Gamma_1}
 \Delta_{\Gamma_2}  \Delta_{\Gamma_3} 
\Delta_{\Gamma_4} .\label{eq:Landau}
\end{equation}
\end{widetext}

\begin{figure}
\includegraphics[scale=.6]{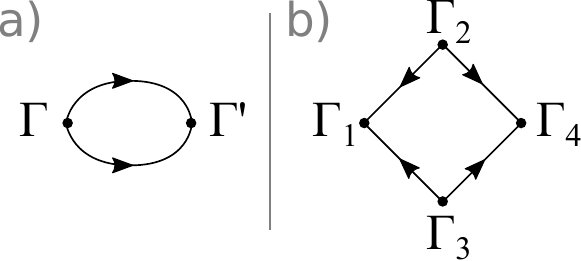} 
\caption{Diagrammatic representation of the matrix pair susceptibility $\chi_{\Gamma\Gamma'}$ [panel a)] and the coefficients $\beta_{\Gamma_1 \Gamma_2 \Gamma_3 \Gamma_4}$ [panel b)] which enter the Landau free energy, Eq.~\eqref{eq:Landau}.}
\label{fig:Susceptibility}
\end{figure}

{H}ere we use {the notation} $\Gamma {\rightarrow C \in \lbrace 1,2\rbrace}$ 
to denote the {two} 
channels {with} repulsive Coulomb {interaction $U_C$. On the other hand, we use $\Gamma \rightarrow P = 3,4,\dots \mathcal D$} to
denote the attractive pairing channels with interaction $g_P$, {for example $s_\pm$ pairing and orbital entangled pairing}. {In Eq.~\eqref{eq:Landau}} a summation over the repeated indices $C,C'$,
$P$ and  $P'$ is implied and {in the limit}
$U_{C}\rightarrow \infty
$, {we can} omit the $1/U_{C}$ terms in the diagonal.
The pair susceptibility $\chi_{\Gamma\Gamma'}$
is a matrix in the space of channels, illustrated diagrammatically
in Fig.~\ref{fig:Susceptibility} along with the coefficients
$\beta_{\Gamma_1 \Gamma_2 \Gamma_3 \Gamma_4}$ of fourth order terms.
Generically, the susceptibilities are finite,
unless $\Gamma$ and $\Gamma'$ belong to different representations of
the tetragonal group. However, sometimes there can be approximate
cancellations brought on by cancellations between electron and hole
pockets, as in the case of the $s^{\pm }$ pairing scenario. 
The mean-field superconducting transition occurs at the first
temperature where the first quadratic matrix develops a zero
eigenvalue, corresponding to the condition
\begin{align}
&0= {\rm det}\left (\begin{array}{cc}
\chi_{CC'} & \chi_{CP'} \\ 
 \chi_{PC'} & -\frac{\delta_{PP'}}{g_P} + \chi_{PP'}
\end{array} \right ) \notag \\
&{ = {\rm det}[\chi_{CC'}]{\rm det}\left[\chi_{PP'}- \chi_{PC}[\chi_{CC'}]^{-1}\chi_{C'P'} -\frac{\delta_{PP'}}{g_P} \right].} \label{eq:TransTemp}
\end{align}
{The term including $[\chi_{CC'}]^{-1}$, which denotes the partial inverse in the subspace of repulsive channels, reduces the susceptibility $\chi_{PP'}$ of pairing channels and thereby suppresses $T_c$.}

{To illustrate these ideas, we analyze a two band model and compare to the discussion of Sec.~\ref{sec:RG}.} We {restrict} our attention to {the} Coulomb channel $C=1$ and $s^{\pm
}$ pairing channel ${P}=3$. In this case, the diagonal
components of the pair susceptibility are
\begin{equation}\label{}
\chi_{CC}\sim \chi_{PP}\sim \rho \ln  \left(\frac{D}{T} \right)
\end{equation}
where as the off-diagonal component is given approximately by
\begin{equation}\label{}
\chi_{CP} \sim \sum_{n=e,h}\int \frac{d^{2}k}{(2\pi)^{2}}  c_{x}c_{y}
\frac{\tanh \frac{\epsilon_{\vk n}}{2{T}}}{2\epsilon_{\vk n}} \sim  \delta \rho \ln \left(\frac{D}{T} \right)
\end{equation}
corresponding to a Fermi surface sum over the electron and hole
pockets. Here, $\delta $ is a parameter that measures the average of
$c_{x}c_{y} { \equiv \cos(k_x) \cos(k_y)}$ 
over the two compensating Fermi surface pockets, weighted
by the density of states. This 
quantity vanishes when the density of states of the 
electron and hole pockets are fully compensated, but it grows to a
number of order unity when either the electron or hole pockets are shrunk to
zero. (Note that in a 
2D band with strict quadratic dispersion,
the density of states is independent of Fermi surface size, and in
this extreme case $\delta $ only becomes sizeable at the Lifshitz
transition
where one Fermi surface vanishes.)

The condition for $T_{c}$ is then given by 
\begin{eqnarray}\label{l}
0&=&{\rm det}\pmat{
\rho  \ln \left(\frac{D}{T_{c}} \right) 
&
\delta \rho  \ln \left(\frac{D}{T_{c}} \right) \cr 
\delta \rho  \ln \left(\frac{D}{T_{c}} \right) & \rho  \ln
\left(\frac{D}{T_{c}} \right)  - \frac{1}{g}
}\cr
&=& \rho \ln \left(\frac{D}{T{_c}} \right) \left[\rho \ln
\left(\frac{D}{T{_c}} \right) (1-\delta^{2})- \frac{1}{g}
\right]
\end{eqnarray}
which gives
\begin{equation}\label{eq:TcUncomp}
T_{c} = D \exp \left[-\frac{1}{g\rho (1-\delta^{2})} \right].
\end{equation}
Thus once the cancellation between the electron and hole pockets is
removed, the effect of the Coulomb interaction is to suppress the
effective coupling constant $g\rho \rightarrow g\rho (1-\delta^{2})$,
producing an exponential depression in the transition temperature. 
Yet experimentally, the elimination of the hole pockets in {iron selenides}
produces no significant reduction in $T_{c}$. 

{Note that this same result ~\eqref{eq:TcUncomp} can also be obtained
using RG method~\eqref{eq:RGeqs} with starting value $g_-(0)$ given
in Eq.~\eqref{eq:gminusbare}. However, the physical discrepancy
between the implications of Eq.~\eqref{eq:TcUncomp} and experimental
reality is independent of the theoretical approach.} This prompts us
to enquire whether the re-introduction of additional orbitals and
pairing channels will provide an additional flexibility to avoid the
strong dependence of $T_{c}$ on Fermi surface structure seen in this
simple example.

We conclude this section with a comment on the incipient band scenario: It has been demonstrated, that bands which do not cross the Fermi level but reside within a distance $D$ in energy space facilitate superconductivity~\cite{HirschfeldMazin2011,Bang2014}. Formally, this is because also incipient bands may contribute large logarithms to $\chi_{P,P'}$. At the same time, the incipient band scenario can not provide a generic reason why the term including $[\chi_{CC'}]^{-1}$ in Eq.~\eqref{eq:TransTemp} should be generically negligible in FeSC and can not solve the Coulomb problem generically.

\subsection{Landau formulation of orbital and k-space flexibility}

The Landau energy has the following salient features: 
First, the quadratic part of the attractive channels is determined by a matrix 
\begin{equation}
[\underline{\epsilon}(T)]_{PP'} =  \frac{\delta_{PP'}}{g_P} - {\chi}_{PP'}, \label{eq:LandauMicro}
\end{equation}
which is positive definite for $T>T_c$ and has lowest eigenvalue proportional to $(T-T_c)/T_c^2$, its associated eigenvector $\hat e$ determining the direction of the instability in channel space. 
Second, in order to avoid the energetically cost of Coulomb repulsion,
{the on-site pair density, i.e. Gor'kov function, must
vanish. Expanding $F_{m,m'}(\v k)$ in s-wave channels,
Eq.~\eqref{eq:VanishingPairCorrelator} becomes $F_C = \chi_{CC'}
\Delta_{C'} + \chi_{CP} \Delta_P =0$ with $C = 1, 2$. The solution
$\Delta_C = - [\chi_{CC'}]^{-1} \chi_{C'P} \Delta_P$ implies a
reduction of the $\chi_{PP'}$ susceptibility, see also}
Eq.~\eqref{eq:TransTemp}, {unless }
the vector $\vec \Delta_{P} = (\Delta_{3}, \Delta_{\rm 4}, \dots)$ of
superconducting order parameters {is} orthogonal to the vectors
formed from the inter-channel pair susceptibilities
 $\vec{\chi }_{C}\equiv \chi_{CP}= (\chi_{C3},\chi_{C4},\dots )$
 (C=1,2). {Consequently, $T_c$ is reduced unless}
\begin{equation}
0 = \vec{\chi }_{C}\cdot \vec{\Delta }_{P}. \label{eq:CoulombConstraint_intro}
\end{equation}

We shall make two assumptions about the vectors
$\chi_{\text{1} P}$,$\chi_{\text{2} P}$ and the matrix $\underline
\epsilon(T)$. First, both vectors and $\underline \epsilon(T)$ are
highly susceptible to changes in the electronic structure, in
particular the Fermi surface geometry. Second, $\chi_{\text{1}
P}=0=\chi_{\text{2} P}$ if the order parameter in pairing channel $P$ transforms according to a different representation than s-wave even-frequency (e.g. if $P$ was a d-wave channel). {In this case Eq.~\eqref{eq:CoulombConstraint_intro} is automatically fulfilled. In contrast, we here concentrate on the case in which $\chi_{CP} \neq 0$ in general, scrutinize another explanation, i.e.} orbital and k-space flexibility, {and therefore ask the following questions}:  does
the large dimension $\mathcal
D$ of the order parameter space 
provide a credible mechanism for the 
FeSC to adapt to the strong on-site Coulomb repulsion? Moreover, 
is this mechanism applicable within the subspace of extended s-wave states?

The appeal of the OKF concept, lies in the observation that 
while there may be several attractive pairing channels, 
there are only two Coulomb constraints
Eq.~\eqref{eq:CoulombConstraint_intro}. 
Although this lowers the dimension of the manifold of attractive
states from $\mathcal D$ to $\mathcal D-2$, 
so long as a Cooper instability survives in one of the many pairing channels
which remain, it would seem plausible that the pairing state can smoothly adapt to the
Coulomb repulsion. The
hope is that this mechanism prevents or at least weakens the
reduction of $T_c$ due to Coulomb repulsion. 

{In order to further develop this idea we first consider the case in which}
the 
lowest eigenvalue of the matrix $\underline \epsilon(T)$ 
{is} almost degenerate, i.e. 
just below $T_c$ there are a number of nearby superconducting instabilities corresponding to a sequence
of small eigenvalues proportional to $(T-T_c^{(2)}),(T-T_c^{(3)}),
\dots$ with $T_c - T_c^{(2,3, \dots)} \ll T_c$. If the
degeneracy is sufficiently large, changes to $T_c$ will remain small
even though the vector space of allowed states perpendicular to
$\chi_{{1 P}}$, $\chi_{{2 P}}$ would 
undergo a large-scale rotation when the electronic structure changes from compound to compound. 

The above discussion demonstrates the {potential} of the OKF
mechanism to overcome the Coulomb repulsion by exploitating 
the large manifold of degenerate pairing states (see
Fig.~\ref{fig:Flexibility}, upper panels). In
particular, the orbital degrees of freedom increase the flexibility by
introducing orbital selective and orbital entangled Cooper channels
which allow to orthogonolize against Coulomb interaction in orbital
space.

\section{Symmetries and Microscopics}
\label{sec:Microscopics}

{In the previous section, the concept of OKF was qualitatively elaborated on the level of a phenomenological Landau theory. In this section, we provide quantitative aspects regarding the large multitude of pairing states, i.e. the fundamental ingredient for OKF, in iron-based superconductors.}
We will focus on the three $t_{2g}$-orbitals {in a purely 2D model}, though our discussion can be simply extended to a more general {setup}.
{We first provide a symmetry classification of states and then present a microscopic calculation.}

\subsection{Superconducting states and crystal symmetries}

\begin{figure}
\includegraphics[scale=.18]{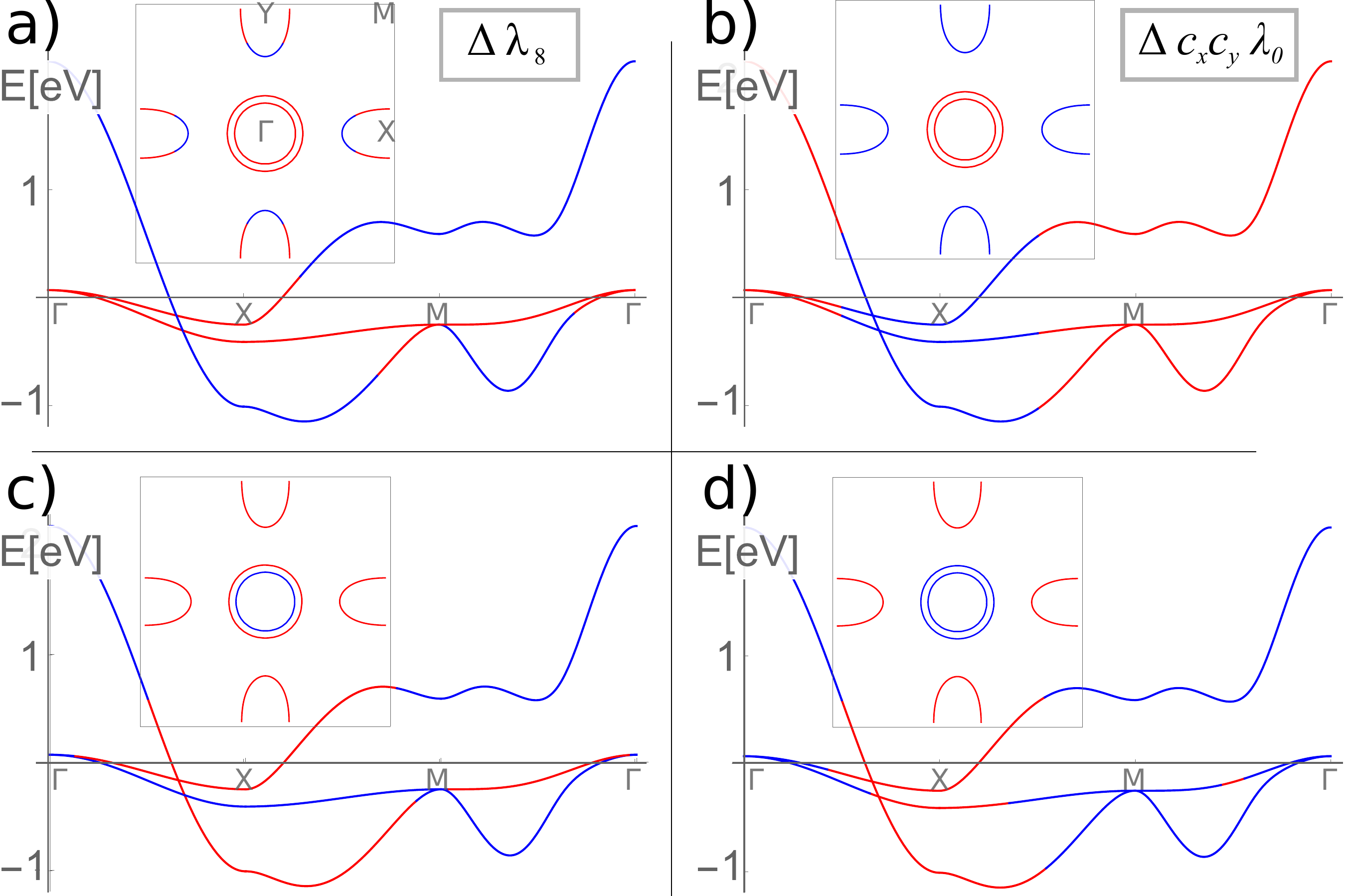} 
\caption{Graphical representation of several superconducting states based on the tight binding model proposed in Ref.~\cite{DaghoferDagotto2010}. The dispersion relation along $\Gamma-X-M-\Gamma$ is colored red (blue) for positive (negative) sign of the gap function. Each panel also contains an inset where the Fermi surfaces are plotted with the same color coding. Panel a): Orbital antiphase, $\Delta (\v k) = \Delta \lambda_8$ (in the present tight binding model, this state has nodes). Panel b): Conventional $s_{\pm}$, $\Delta (\v k) = \Delta c_x c_y \lambda_0 $. Panel c): A mixed state $\Delta (\v k) = \Delta (-0.01c_x c_y \lambda_0  - 0.99 s_x s_y \lambda_1 +0.11 (c_x - c_y) \lambda_3)$ with strong orbital entanglement. Panel d): A mixed state $\Delta (\v k) = \Delta (-0.97 c_x c_y \lambda_0  - 0.03 s_x s_y \lambda_1 +0.24 (c_x - c_y) \lambda_3)$ which is mainly s$_\pm$. Panel c) [d)] occurs at $x = -0.9$ [$x = 0.7$] in the microscopic calculation of Fig.~\ref{fig:xDependence}. Note that orbital entanglement naturally explains the sign change within the two inner Fermi surfaces of hole doped FeSC as proposed in~\cite{MaitiChubukov2013}. }
\label{fig:SuperconductingStates}
\end{figure}

The matrix gap function $\Delta (\vk) $ can naturally be
expanded in irreducible representations of the point group
\begin{equation}
\Delta_{m m'} (\vk)  = \sum_{\Gamma = 1}^{\mathcal D}
\Delta_\Gamma \varphi^{\Gamma}(\v k) [\lambda_{i(\Gamma)}]_{mm'}.\label{eq:ExpansionChannels}
\end{equation}
Here, $\lambda_i$ are the Gell-Mann matrices (see Appendix
\ref{app:GellMann}).
A comprehensive
classification ~\cite{Fischer2013} of all states based on the
crystal symmetries is summarized {in Table ~\ref{tab:Pairingstates}}, with more details given 
in appendix~\ref{app:ClassificationOfStates} and an illustration presented in Fig.~\ref{fig:SuperconductingStates}. The coefficients
$\Delta_\Gamma$ are real, provided time-reversal symmetry is
unbroken. For 
spin singlet pairs, the exclusion principle forces the spatial wavefunction to be even, 
so that the \textit{combined} form factors
$\varphi^{\Gamma}(\v k) \lambda_{i(\Gamma)}$ are even
under simultaneous spatial inversion and matrix transposition. 
Point group operations, such as 
a $\pi$ rotation in the iron plane $\v k \rightarrow R_\pi \v k =
(-k_x,-k_y, k_z)$, are represented by unitary matrices in orbital
space, e.g. $U_{R_\pi} = \text{diag}(-1,-1,1)$. We adopt a convention
in which the orbitals are placed in the order
$m= (1,2,3)\equiv (d_{xz},d_{yz}, d_{xy})$. Assuming the order
parameter has overall s-wave symmetry, it must remain invariant under the
crystal transformations, 
\begin{equation}
\Delta (\v k) = U^{\dagger}_{R_\pi} \Delta (R_{\pi} \v k) U_{R_\pi}. \label{eq:TrafoMainText}
\end{equation}
Still, non-trivial (e.g. $B_{1g}$) form factors $\varphi^{\Gamma}(\v k)$ are possible, as long as the transformation behavior of associated Gell-Mann matrices $\lambda_{i(\Gamma)}$ compensates the transformation behavior of $\varphi^{\Gamma}(\v k)$. 

In {Table ~\ref{tab:Pairingstates}} we stretch our notation beyond standard group theoretical convention and use different $\Gamma$ indices for form factors $\varphi^{\Gamma}(\v k)$ of the same irreducible representation, but with different Fourier harmonics. This allows to distinguish physically distinct states with the same transformation properties, e.g. conventional $s$-wave from the $s_\pm$ {or orbital entangled} states.
 This 
 description
can be extended in various ways, e.g to include higher angular
momentum or odd-frequency pairing ~\cite{LinderBalatsky2017}.

\subsection{Derivation of Landau theory}
\label{sec:LandauTheory}

A BCS description of the interactions in these various 
channels then takes the form 
\begin{equation}\label{}
H_{I} =  \sum_{
C = 1,2}
U_{C}
\Psi \dg_{C}\Psi_{C} 
- \sum_{P = 3, \dots}  
g_{{P}} 
\Psi \dg_{{P} }\Psi_{{P}},
\end{equation}
where we use $C\equiv \Gamma_{C}$ 
to denote the repulsive Coulomb channels and $P\equiv \Gamma_{C}$ to
denote the attractive pairing channels. The 
pair creation operators take the form 
\begin{equation}\label{}
\Psi \dg_{\Gamma }= \sum_{\vk, m,m' }\varphi^{\Gamma} (\vk )\lambda^{\Gamma}_{m,m'}
\left(
c\dg_{\vk m \uparrow} 
c\dg_{-\vk m'\downarrow } - (\uparrow \leftrightarrow \downarrow )\right),
\end{equation}
where the sum is over an energy shell
$ |\epsilon_{\vk }| \leq {D} $ around the Fermi surface.

We typically
carry out a Hubbard-Stratonovich transformation of the repulsive and
attractive interactions, so that the interactions can be written in a
mean-field form 
\begin{equation}\label{l}
H_{I} = \sum_{\Gamma}\left(\Delta_{\Gamma} \Psi_{\Gamma }+ {\rm H.c}
\right)
- \sum_{C{= 1,2}}\frac{|\Delta_{C}|^{2}}{U_{C}}
+ \sum_{P= 3, \dots }\frac{|\Delta_{P}|^{2}}{g_{P}}.
\end{equation}
This kind of transformation is carried out with the 
understanding that inside the path integral, the Coulomb gap
variables are integrated along the imaginary axis $\Delta_{C}\in
[-i \infty, i \infty ]$, but that in a mean-field theory, we can
distort the contour and seek a saddle point solution 
where the $\Delta_{C}$ lie on the real axis. 
The combined set of order parameters 
$\Delta_\Gamma = (\Delta_{C},\Delta_{P})
$
then form a 
$\mathcal D$ dimensional real vector.

We are particularly interested in the limit where the
$U_{C}\rightarrow \infty $ are large enough to neglect the second term. In this case, the saddle
point values of the 
$\Delta_{C}$ (C=1,2) play the role of Lagrange multipliers that impose
the Coulomb constraints  $\langle \Psi_{C}\rangle =0$ (C=1,2), equivalent to
conditions (\ref{localcons}).

{The derivation of the Landau free energy follows from the integration of fermionic degrees of freedom.}
The generic form for the pair susceptibility matrix {entering Eq.~\eqref{eq:Landau}} is
given by 
\begin{equation}\label{}
\chi_{\Gamma\Gamma'}=\sum_{n,n'} \int \frac{d^{2}k}{(2\pi)^{2}}  \varphi_{n,n'}^{\Gamma} (\vk )
\varphi^{\Gamma'}_{n',n} (\vk )
\frac{1 - f (\epsilon_{\vk n})-f (\epsilon_{\vk n'})}{\epsilon_{\vk n}+\epsilon_{\vk n'}}.
\end{equation}
where $n$ is the band-index and the 
\begin{equation}\label{}
\varphi_{n,n'}^{\Gamma} (\vk ) = \varphi^{\Gamma} (\bk )\sum_{m,m'} \langle \vk n|m\rangle 
\lambda^{\Gamma}_{m,m'}\langle m'\vert \vk  n\rangle 
\end{equation}
are the form-factors of the pairing matrices in the band-basis. The
interband $(n\neq n')$ parts of these matrices are weakly temperature
dependent. However, the intra-band parts $(n=n')$ contain a
Cooper-instability divergence
\begin{equation}\label{}
\chi_{\Gamma\Gamma'} (T) = \rho_{\Gamma \Gamma'}
\ln \frac{D}{T}
\end{equation}
where 
\begin{equation}\label{}
\rho_{\Gamma\Gamma'}= \sum_{n}\rho_{n}\left\langle \varphi^{\Gamma}_{nn}
(\vk ) \varphi^{\Gamma'}_{nn} (\vk )\right\rangle_{FS} 
\end{equation}
where $\rho_{n}$ is the density of states 
for the n-th Fermi surface
and $\langle \dots \rangle_{FS}$ represents the corresponding Fermi
surface average. 

\subsection{Generalized t-J model}
\label{sec:tJ}

\begin{figure}
\includegraphics[scale=.38]{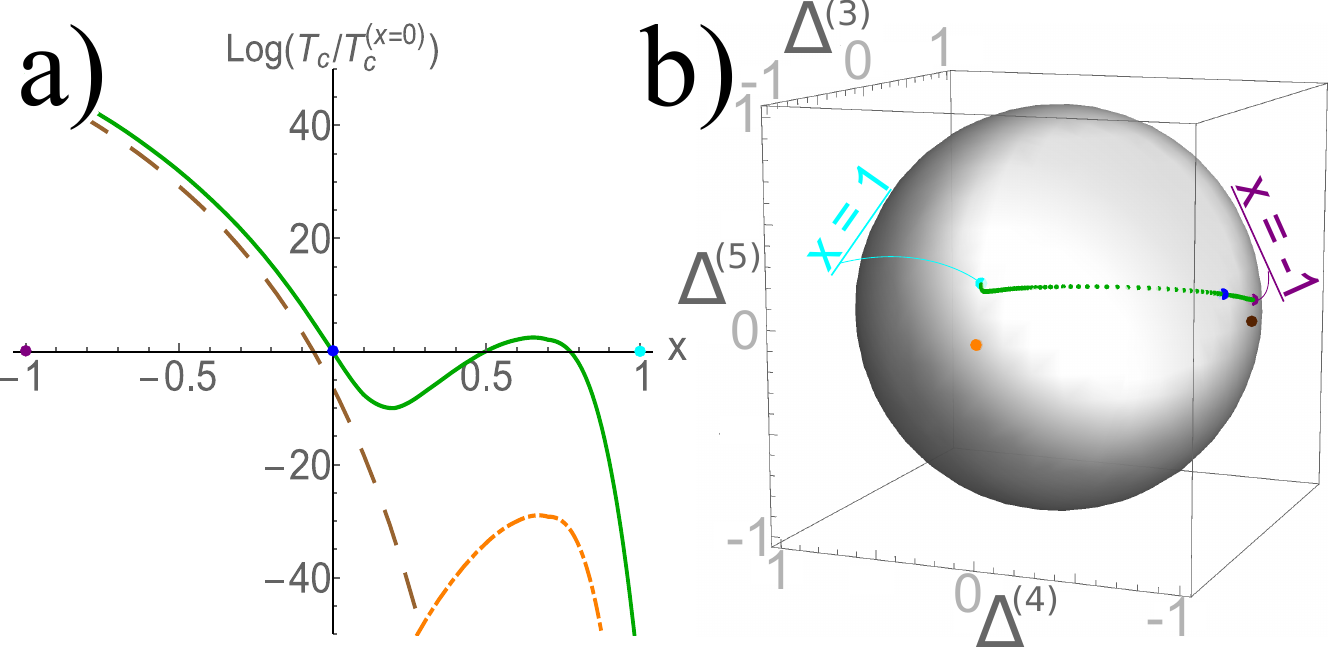} 
\caption{Microscopic investigation of orbital flexibility based on Eqs.~\eqref{eq:Landau},~\eqref{eq:LandauIAs} of the main text. The parameter $x$ is introduced in the density of states $\rho_e = (1+x) /(20J)$ ($\rho_h = (1-x) /(10J)$) of electron pockets (hole pockets) {in order to smoothly interpolate between systems with different types of carriers}. Panel a): Evolution of $T_c$ as a function of $x$, the solid/dashed/dotdashed curves represent $T_c$ for the flexible/the $\Delta_{4}$/the $\Delta_{3}$ state, all of which are normalized to the $T_c$ of the flexible state at $x = 0$. Panel b): Evolution of the direction of the flexible state in the space of $\Delta_{3,4,5}$. For this plot we chose a moderate $\bar U/J = 3$. For $\bar U/J = 10$ the graphs of both panels are qualitatively similar, however the local maximum near $x \approx 0.7$ drops down to $\ln(T_c/T_c^{(x = 0)}) \approx -20$.   }
\label{fig:xDependence}
\end{figure}

In this section we {investigate} our findings by means of a microscopic t-J model (without a constraint on the occupancy) treated to the mean field level. In addition to the on-site interaction \eqref{eq:HKanamori} we consider nearest neighbor antiferromagnetic coupling $J_1$ and next nearest neighbor antiferromagnetic coupling $J_2$,
\begin{subequations}
\begin{align}
H_{\rm J_1} &= \sum_{m m'} \sum_{\v k}  J_1^{m m'} (c_x+c_y) :\v S_{m, \v k} \cdot \v S_{m', - \v k } :,\\
H_{\rm J_2} &=\sum_{m m'} \sum_{\v k}  J_2^{m m'} c_xc_y :\v S_{m, \v k} \cdot \v S_{m', - \v k }:.
\end{align}
\label{eq:Interactions2}
\end{subequations}
{W}e note that inter-orbital magnetic interactions are not uncommon, 
{e.g.}~perfectly Hund's aligned 
antiferromagnetic interaction can be expected to be of the form $J \v S_{\rm total,1} \cdot \v S_{\rm total,2}  = J (\sum_m \v S_{m,1}) \cdot (\sum_{m'} \v S_{m',2})$ projected onto the strong Hund's subspace. 

In Appendix~\ref{app:tJIA}, we project the interactions on the space of spin singlet pairing. The magnetic interactions generate Cooper channel attraction in extended s-wave 
and, for concreteness, {we} assume that Cooper instabilities occur only in channels $\Gamma = {3 - 5}$, so that 
the effective Landau free energy, \eqref{eq:Landau}, is determined by only three attractive interactions 
\begin{subequations}
\begin{eqnarray}
g_{{3}}&=& [2 J_{2}^{xz,xz}+J_{2}^{xy,xy}]/4, \\
g_{{4}}&=& [3 J_{2}^{xz,yz} ]/4, \\
g_{{5}}&=& [3J_{1}^{xz,xz}]/8,
\end{eqnarray} 
\label{eq:LandauIAs}
\end{subequations}
{and repulsions $U_1 = U_2 = [U + 3J_{H,1}/2]/2 \equiv \bar U/2$.} We used $J_{2}^{xz,xz} = J_{2}^{yz,yz}$ due to crystal symmetries.

In Fig.~\ref{fig:xDependence} we numerically evaluate $T_c$ for the given system and introduce a parameter $x \in (-1,1)$ to interpolate between the hole dominated case at $x = -1$ to an electron dominated case at $x = 1$. 
We assume $[2 J_{2}^{xz,xz}+J_{2}^{xy,xy}]/3 =J_{2}^{xz,yz} = 2 J_{1}^{xz,xz} \equiv J/2$ {and}
Fermi surfaces with the orbital content and geometry analogous to those~\cite{DaghoferDagotto2010} represented in Fig.~\ref{fig:SuperconductingStates}. 
We compare our results to the fictitious case of a system in which orbital and k-space flexibility are not allowed. Clearly, the system takes advantage of its flexibility, the instability changes direction from $(\Delta_{3}, \Delta_{4}, \Delta_{5})\propto (0, -0.99, 0.11)$ at $x = -1$ to $(\Delta_{3}, \Delta_{4}, \Delta_{5})\propto (-0.97, -0.03, 0.24)$ at $x = 0.7$, the position of optimal electron doping, see Fig.~\ref{fig:SuperconductingStates} c) and d) for an illustration of these {orbital entangled and orbital selective} states. 
Panel a) of Fig~\ref{fig:xDependence} demonstrates how the system adapts from following $T_c^{(4)}$ in channel $\Delta_{4}$ to the $T_c^{(3)}$ of $\Delta_{3}$ as soon as the orange dotdashed curve overtakes the brown dashed curve.
However, Fig.~\ref{fig:xDependence} also demonstrates, that the maximum $T_c^{(3)}$ and the maximum $T_c^{(4)}$ are generically well separated, even for the present choice of equal attractive interaction in the two channels. This results in $\ln[T_c^{(x = - 1)}/T_c^{(x = 0.7)}] \sim 40$ for the present choice $\bar U/J = 3$, which is gigantic and increases further as $\bar U/J$ increases. {These observations are the basis for the criticism of OKF presented in the next section.}

\section{Critique and Outlook}
\label{sec:SummaryOutlook}

A key result of our investigation of 
the effects
of onsite Coulomb interactions in the iron-based 
superconductors, is that observation that 
Coulomb repulsion enforces two constraints on the
condensate, forcing two 
on-site, equal-time pairing amplitudes to vanish, {see Eq.~\eqref{eq:CoulombConstraint_intro}}. In most
unconventional superconductors, these Coulomb constraints  are
symmetry-protected 
 through the development of higher angular momentum
condensates, for which the Coulomb constraint is automatically
satisfied.  Current extended s-wave theories of the iron-based superconductors
have difficulty explaining how the Coulomb constraint is satisfied for
a wide variety of Fermi surface morphologies, without marked
suppression of $T_{c}$. {A key question is whether there is some kind
of  hidden or accidental symmetry in FeSC?}

We attempted to provide a way out of this 
conundrum, examining a hypothesis that pairing within a large class
of orbitally entangled singlet states allows the condensate the
orbital and k-space flexibility (OKF) to ``rotate'' into new configurations
that satisfy the Coulomb constraint in new Fermi surface morphologies.
We encountered a number of unresolved issues with the OKF approach:
\begin{enumerate}
\item OKF requires that the
lowest eigenvalues of the matrix $\underline \epsilon(T)$ must
be almost degenerate, see Sec.~\ref{sec:PhenomenologicalLandau}.~\footnote{If
the lowest eigenvalue was not degenerate, one would be forced to
require the eigenvector $\hat e$ to change along with $\chi_{\text{I}
C}$,$\chi_{\text{II} C}$ as the electronic structure alters. Such a
mechanism is unknown and, in any case, different from the proposed
orbital and k-space flexibility.} 

\item From a microscopic weak-coupling viewpoint, 
different Fermi surface geometries lead to different pair
susceptibility matrices. Therefore, even if the lowest eigenvalue of
$\underline{\epsilon}(T)$ was largely degenerate for a given compound,
it is unlikely to be so in another. {This would lead to dramatic
variations in $T_c$ from compound to compound, see
Fig.~\ref{fig:Flexibility}, lower panels.}

\item Concerning
k-space flexibility, one would like to concentrate on states involving
only the first few harmonics, since higher harmonics require 
long-distance interactions. This substantially reduces the
dimension $\mathcal D$ of the order parameter space. Finally,
it is worth mentioning that a linear increase in the lowest
eigenvalue of $\underline \epsilon$ implies an exponential decrease of
$T_c$ within microscopic weak-coupling theories, due to the
logarithmic Cooper instability.
\end{enumerate}
These
arguments
suggest that orbital and k-space
flexibility 
is unable to explain
the robustness of $T_c$ against Coulomb repulsion in the zoo of
FeSC. {Our microscopic calculations leading to Fig.~\ref{fig:xDependence} corroborate this conclusion.} The OKF mechanism requires a well-protected degeneracy of
transition temperatures, which in turn would require an underlying
symmetry which is simply absent for an orbitally entangled s-wave condensate.

In the present analysis, only Cooper channel instabilities were
considered in a simple mean field treatment. More generally
particle-hole channels should also be taken into account. As we mentioned
in the introduction {and in Sec.~\ref{sec:RG}}, while the mutual renormalizations of particle-hole and
particle-particle channels can provide a mechanism to reverse the
sign of Coulomb {interaction} near the scale of the superconducting
instability, the results are still highly sensitive to the Fermi
surface morphology and the pattern of bare repulsive interaction values
~\cite{KoenigColeman2017}. In short, the existing weak-coupling RG scenarios are 
unable to account for the robustness against changes of Fermi surface morphologies as $T_{c}$ strongly depends on ultraviolet scales.

What alternative mechanisms for robustly avoiding the Coulomb problem
might be at work across the full range of iron-based superconductors? 
One possibility we are forced to reconsider, 
is that pairing in the FeSC is not primarily driven by 
extended s-wave, spin singlet pairing, but that 
it involves a symmetry-protected orthogonalization against
Coulomb interaction.  One way forward is to 
explore higher angular momentum states in the spin singlet
channel. Many proposals in the literature have examined 
d-wave~\cite{KhodasChubukov2012,ChubukovFernandes2016,NicaSi2017,AgterbergWeinert2017}
pairing. 

The common observation of full
excitation gaps remains an open issue for higher angular momentum
states (see Ref.~\cite{ChubukovFernandes2016} for a thorough
discussion of the disappearance of nodes).  A second way forward is to
consider spin triplet pairing. Apart from early p-wave proposals
~\cite{LeeWen2008}, the orbital degrees of freedom  allow for the
formation of condensates which are antisymmetric in orbital
space. Such ``orbital singlet'', spin triplet
pairing~\cite{VafekChubukov2017,CheungAgterberg2018} can occur in an s-wave channel. 
The classic objection to spin triplet pairing is the observation of a Knight
shift. However,  spin orbit coupling can lead to spin singlet
admixture. Indeed, the Knight shift in the spin triplet
superfluid $^3$He is actually sizeable, in part because of spin-orbit coupling
~\cite{VollhardtWoelfle1990}.  Finally, odd-frequency pairing might
find another route to  
orthogonalization against the Coulomb constraints. While such a state
state does not develop 
a spectral gap on its own, it may couple to pairing
channels in which the gap matrix and the kinetic part Hamiltonian do
not commute and thus indirectly induce a gap.

\section{Acknowledgements}
We acknowledge valuable discussions with T. Ayral, P.-Y. Chang and
Y. Komijani. This material is based upon work supported by the U.S. Department of Energy, Office of Science, Office of Basic Energy Sciences, under Award DE-FG02-99ER45790 (Elio Koenig and Piers Coleman).
\appendix

\section{Classification of the order parameter matrix within the three band model}
\label{app:ClassificationOfStates}

This Appendix reviews the crystal symmetries of
iron based superconductors and deduces their implications
for the kinetic energy and pairing components of the Hamiltonian
within the space of  $t_{2g}$ orbitals. This provides us with a comprehensive
classification of superconducting states within the three-orbital
model. This work extends earlier work~\cite{Fischer2013} to the case of
band-off-diagonal gap functions, which in turn
implies~\cite{BlackSchafferBalatsky2013} odd frequency pairing.

Consider the following Gor'kov Green's function
\begin{equation}
\mathcal G_{\v k}(\tau) = - \langle T \Psi_{\v k}(\tau) \Psi^\dagger_{\v k}(0) \rangle. \label{eq:GorkovGFtau}
\end{equation}
Here the Nambu spinors $\Psi_{\v k}$ are composed 
from fermionic creation and annihilation operators $c_{\v k, \alpha, m}^\dagger, c_{\v k, \alpha, m}$ in the standard way $\Psi_{\v k} = (c_{\v k, \uparrow}^T, c^\dagger_{-\v k, \downarrow})^T$ and $\Psi^\dagger_{\v k} = (c_{\v k, \uparrow}^\dagger, c_{-\v k, \downarrow}^T)$. The transposition symbol indicates that creation (annihilation) operators are grouped into row (column) three-vectors and we choose to order orbitals $m = d_{xz},d_{yz},d_{xy}$ from left to right (top to bottom). The Gor'kov Green's function is a matrix in Nambu space which, in frequency representation, can be displayed as
\begin{subequations}
\begin{align}
\mathcal G_{\omega_n,\v k}& = \left (\begin{array}{cc}
G^{(e)}_{\omega_n,\v k} & F_{\omega_n,\v k} \\ 
F^\dagger_{-\omega_n,\v k} & G^{(h)}_{\omega_n,\v k}
\end{array} \right ) \label{eq:GorkovGFfreq} \\
&= \Bigg (\begin{array}{cc}
i \omega_n - H_{\v k} - \Sigma_{\omega_n, \v k} & -\Delta_{\omega_n,\v k} \\ 
-\Delta^\dagger_{-\omega_n, \v k} & i \omega_n + H_{-\v k}^T + \Sigma_{-\omega_n, -\v k}^T
\end{array} \Bigg )^{-1} . \label{eq:GorkovGFMF}
\end{align}
\label{eq:GorkovGFepsFull}
\end{subequations}
All entries of these matrices are themselves three by three matrices in orbital space. We used $\mathcal G_{\omega_n,\v k}  = \mathcal G^\dagger _{- \omega_n, \v k} $, cf. the definition \eqref{eq:GorkovGFtau}, so that electron and hole Green's functions satisfy $G^{(e,h)}_{\omega_n,\v k}= [G^{(e,h)}_{-\omega_n, \v k}]^\dagger$. In the second line, we relate the Gor'kov Green's function to the kinetic part of the Hamiltonian $H_{\v k}$, a self-energy $\Sigma_{\omega_n,\v k}= \Sigma_{-\omega_n, \v k}^\dagger$ and a gap function $\Delta_{\omega_n, \v k}$. In full generality, the order parameter $F$ and gap $\Delta$ matrices may be frequency dependent, for a comment on the subtlety of the reversed frequency in the lower left matrix element see Sec.~\ref{sec:OddFreq}. 

\subsection{Symmetry operations}

\begin{table}
\begin{tabular}{|c|c|c|c|| l |}
\hline 
$\alpha_1 \leftrightarrow \alpha_2$ & $\omega_n \leftrightarrow - \omega_n$ & $\v k \leftrightarrow - \v k$ & $m_1 \leftrightarrow m_2$ & Details \\
\hline \hline
$-$ & + &+&+& Table~\ref{tab:ppp} \\
\hline
$-$ & + &$-$&$-$& Table~\ref{tab:pmm} \\
\hline
$-$ & $-$ &+&$-$& Table~\ref{tab:mpm} \\
\hline
$-$ & $-$ &$-$&+& Table~\ref{tab:mmp} \\
\hline 
\end{tabular} 
\caption{Summary of superconducting singlet states for the three-orbital model of the iron-based superconductors. The first column refers to the parity under exchange of the spins of the constituent electrons of the Cooper pairs. The second and third column refer to the parity under temporal or spatial inversion of the order parameter function. The fourth column describes the parity under transposition in the space of $t_{2g}$ orbitals. By assumption, the order parameter field is in a singlet state and thus odd under exchange of spin. For each row, Pauli's principle imposes a negative sign upon multiplication of columns one through four. The last column refers to tables~\ref{tab:ppp} - \ref{tab:mmp} containing more details on each of the four category of states.}
\label{tab:summary}
\end{table}

\begin{figure}[tbh]
\includegraphics[width=.8\columnwidth]{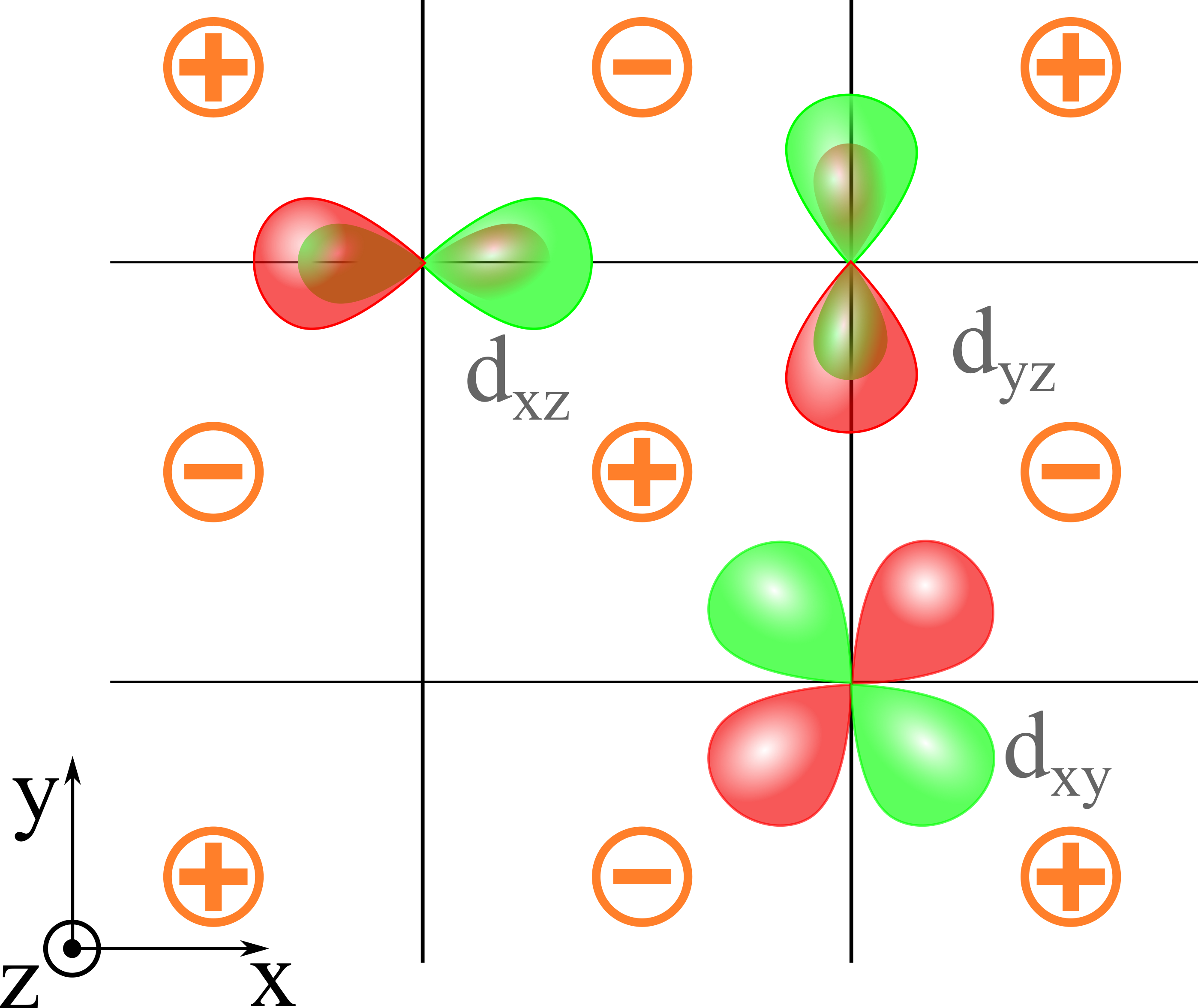} 
\caption{Representation of the three $t_{2g}$ orbitals in an iron plane. The sites of the lattice correspond to iron positions. Chalcogen/pnictogen atoms reside in alternating positions above/below a plaquette (indicated by a circle with a plus/minus sign). The sign of the lobes of the orbitals is represented by two different colors: light green and darker red.}
\label{fig:Orbitals}
\end{figure}

As we mentioned in the main text, we are interested in singlet
pairing. The Pauli priniciple imposes (see also Table ~\ref{tab:summary})
\begin{equation}
F_{\omega_n, \v k} =   F_{- \omega_n,-\v k}^T,\quad \Delta_{\omega_n,\v k}= \Delta_{- \omega_n,-\v k}^T. \label{eq:Pauli}
\end{equation}
Furthermore, time reversal symmetry implies $\mathcal G_{\omega_n, \v k} = \mathcal G^T_{\omega_n, -\v k}$ so that $G^{(e,h)}_{\omega_n, \v k} =  (G^{(e,h)}_{\omega_n, -\v k})^T$, $H_{\v k} = H_{-\v k}^T$, and $\Sigma_{\omega_n, \v k} =  \Sigma_{\omega_n, -\v k}^T$ as well as hermiticity of the anomalous Green's function and the gap matrix
\begin{equation}
F_{\omega_n, \v k} =  F_{\omega_n, \v k} ^\dagger,\quad \Delta_{\omega_n,\v k}= \Delta_{\omega_n,\v k}^\dagger. \label{eq:TR}
\end{equation}
Since the chalcogen/pnictogen atoms are
ordered above and below the iron plane in an alternating manner, the
system has a two-atom unit cell which in iron-only models shows up as staggered hopping elements. We choose the origin on an iron
site for the lattice model represented by
Fig~\ref{fig:Orbitals}. 

There are two mirror symmetries
\begin{subequations}
\begin{eqnarray}
\mathcal G_{\omega_n,\v k} &=& \left (\begin{array}{ccc}
1 & 0 & 0 \\ 
0 & -1 & 0 \\ 
0 & 0 & -1
\end{array} \right ) \mathcal G_{\omega_n, \hat P_x \hat P_z \v k} \left (\begin{array}{ccc}
1 & 0 & 0 \\ 
0 & -1 & 0 \\ 
0 & 0 & -1
\end{array} \right )\notag 
\\
&=&\left (\begin{array}{ccc}
-1 & 0 & 0 \\ 
0 & 1 & 0 \\ 
0 & 0 & -1
\end{array} \right ) \mathcal G_{\omega_n,\hat P_y \hat P_z \v k} \left (\begin{array}{ccc}
-1 & 0 & 0 \\ 
0 & 1 & 0 \\ 
0 & 0 & -1
\end{array} \right ) , 
\end{eqnarray}
\label{eq:Mirrors}
\end{subequations}
and a $C_4$ rotation symmetry
\begin{equation}
\mathcal G_{\omega_n,\v k} = \left (\begin{array}{ccc}
0 & 1 & 0 \\ 
-1 & 0 & 0 \\ 
0 & 0 & -1
\end{array} \right ) \mathcal G_{\omega_n, \hat R_{\pi/2} \hat P_z \v k} \left (\begin{array}{ccc}
0 & -1 & 0 \\ 
1 & 0 & 0 \\ 
0 & 0 & -1
\end{array} \right )  . \label{eq:Rotation}
\end{equation}
Here we have introduced a set of three inversion operators 
$(\hat{P}_{x}, \hat P_{y}, \hat P_{z})$ which respectively reverse 
the x, y and z components of momentum, and a rotation $\hat R_{\pi/2}: (k_x,k_y,k_z)\rightarrow(k_y,-k_x,k_z)$.

\subsection{Even frequency pairing}
\label{app:EvenFreqClass}

Often, superconducting order parameters are assumed to be an even function of frequency (in most cases a constant). We denote such states by a ``$+$'' in the subscript of the order parameter $F_{\omega_n, \v k,+}$ and gap $\Delta_{\omega_n, \v k,+}$. According to Eqs.~\eqref{eq:Pauli},\eqref{eq:TR} these matrix functions transform in the same manner as the kinetic part of the Hamiltonian under crystal and time reversal symmetries. It is convenient to expand the order parameter matrix by means of Gell-Mann matrices (see Appendix~\ref{app:GellMann})
\begin{equation}
F_{\omega_n, \v k,+} = \sum_{i = 0}^8 F^{ (i)}_{\omega_n, \v k,+} \lambda_i . \label{eq:ExpFp}
\end{equation}
The gap function $\Delta_{\omega_n, \v k,+}$ and the diagonal parts of
the (inverse) Green's function, and the Hamiltonian are expanded
analogously. The transformation behavior of each function
$F^{(i)}_{\omega_n, \v k,+}, \Delta^{ (i)}_{\omega_n, \v k,+},$ etc.
may be found in Table ~\ref{tab:ppp} and Table ~\ref{tab:pmm}, corresponding to symmetric and antisymmetric Gell-Mann matrices, respectively. We note that, in a purely $2D$ model where $k_z = 0$, the coefficients of  $\lambda_{2,4,6}$ vanish by symmetry.

\begin{table}
\begin{tabular}{|c||c|c|c|c||c|}
\hline 
$i$ & $\hat P_x$ & $\hat P_y$ &$\hat P_z$ & $\hat R_{\pi/2}$ &examples \\ 
\hline \hline
 0 & +&+ &$+$ &  +  &$1, c_z , c_xc_y$ \\ 
\hline 
1  & $ -$ &$ -$ &$+$ &  $ -$& $s_xs_y$ \\ 
\hline 
3  & + & + &$+$ & $ -$  & $(c_x-c_y)$  \\  
\hline 
$\Bigg ( \begin{array}{c}4\\ 6\end{array}  \Bigg )$ &$ \Bigg ( \hspace{-.1cm} \begin{array}{cc}
+ & 0 \\ 
0 & -
\end{array} \hspace{-.1cm} \Bigg ) $&$ \Bigg (\hspace{-.1cm}\begin{array}{cc}
- & 0 \\ 
0 & +
\end{array}\hspace{-.1cm}\Bigg ) $&$ \Bigg (\hspace{-.1cm}\begin{array}{cc}
- & 0 \\ 
0 & -
\end{array}\hspace{-.1cm}\Bigg ) $&$ \Bigg (\hspace{-.1cm}\begin{array}{cc}
0 & - \\ 
+ & 0
\end{array}\hspace{-.1cm}\Bigg )$& $\Bigg (\hspace{-.1cm}\begin{array}{c} s_y s_z\\ s_x  s_z\end{array}\hspace{-.1cm}\Bigg ) $ \\ 
\hline 
8 & + & + &$+$ & + & $1, c_z, c_xc_y$  \\ 
\hline 
\end{tabular} 
\caption{Transformation behavior of coefficients $F^{ (i)}_{\omega_n, \v k,+}, \Delta^{ (i)}_{\omega_n, \v k,+}, H^{ (i)}_{\v k}$, see Eq.~\eqref{eq:ExpFp}, which are even under all of $\omega_n \rightarrow - \omega_n$, $\v k \rightarrow - \v k$ and transposition in orbital space. The signs indicate the parity under reflections and under $\pi/2$ rotations in the iron plane. The functions $(F^{ (4)}_{\omega_n, \v k,+},F^{ (6)}_{\omega_n, \v k,+})$ (and analogously for $\Delta_{\omega_n, \v k}$ and $H_{\v k}$) transform under a two dimensional representation. These two functions get interchanged under a $\pi/2$ rotation. The last column displays the leading harmonics. }
\label{tab:ppp}
\end{table}

\begin{table}
\begin{tabular}{|c||c|c|c|c||c|}
\hline 
$i $& $\hat P_x$ & $\hat P_y$ &$\hat P_z$ & $\hat R_{\pi/2}$ & examples \\ 
\hline \hline
2 & + & + & $ -$& $ -$ &  $(c_x-c_y)s_z$  \\  
\hline 
$\Bigg (\begin{array}{c}5\\ 7 \end{array}\Bigg )$ &$ \Bigg (\hspace{-.1cm}\begin{array}{cc}
- & 0 \\ 
0 & +
\end{array}\hspace{-.1cm}\Bigg ) $&$ \Bigg (\hspace{-.1cm}\begin{array}{cc}
+ & 0 \\ 
0 & -
\end{array}\hspace{-.1cm}\Bigg ) $&$ \Bigg (\hspace{-.1cm}\begin{array}{cc}
+ & 0 \\ 
0 & +
\end{array}\hspace{-.1cm}\Bigg ) $&$ \Bigg (\hspace{-.1cm}\begin{array}{cc}
0 & + \\ 
- & 0
\end{array}\hspace{-.1cm}\Bigg )$& $\Bigg (\hspace{-.1cm}\begin{array}{c} s_x \\ s_y  \end{array}\hspace{-.1cm}\Bigg )$ \\ 
\hline 
\end{tabular} 
\caption{Transformation behavior of coefficients $F^{(i)}_{\omega_n, \v k,+}, \Delta^{(i)}_{\omega_n, \v k,+}, H^{(i)}_{\v k}$, see Eq.~\eqref{eq:ExpFp}, which are even under $\omega_n \rightarrow - \omega_n$ but odd under $\v k \rightarrow - \v k$ and transposition in orbital space. The notation is the same as in Table~\ref{tab:ppp}.}
\label{tab:pmm}
\end{table}

\subsection{Odd frequency pairing}
\label{sec:OddFreq}
Odd frequency components~\cite{LinderBalatsky2017} of the order parameter and gap matrix, which we denote as $F_{\omega_n, \v k,-}$, $\Delta_{\omega_n, \v k,-}$ are generally considered to be more exotic than even-frequency states. 
However, in multiband materials, odd-frequency components
$F_{\omega_n, \v k,-}$ are also induced by an even $\Delta_{\omega_n,
\v k,+}$ in the case of interband
pairing~\cite{BlackSchafferBalatsky2013}, e.g. for
 $\mathcal G_{\omega_n, \v k} = [i \omega_n -
H_{\v k} \tau_z - \Delta (\v k) \tau_x]^{-1}$, we find
\begin{align}
&\frac{\mathcal G_{\omega_n, \v k}  - \mathcal G_{-\omega_n, \v k} }{2}\simeq- i \omega_n \frac{1}{\omega_n^2 + H_{\v k}^2 + \Delta (\v k)^2 } -\omega_n  \tau_y \times \notag \\
&\times  \frac{1}{\omega_n^2 + H_{\v k}^2 + \Delta (\v k)^2 } [H_{\v k}, \Delta (\v k)]  \frac{1}{\omega_n^2 + H_{\v k}^2 + \Delta (\v k)^2 } . \label{eq:OddFreqFromEvenGap}
\end{align}
A frequency independent
gap function $\Delta (\v k)\tau_x$ thus generates an odd frequency
anomalous Green's function proportional to $\tau_y$. 
Analogously to the even frequency case we classify the components of the pair amplitude
\begin{equation}
F_{\omega_n,\v k,-} = \sum_{i=0}^{8} F^{(i)}_{\omega_n,\v k,-}\lambda_i . \label{eq:ExpFm}
\end{equation}
The transformation behavior of all components is presented in tables~\ref{tab:mpm} and \ref{tab:mmp}, where the former (latter) table is devoted to states which are even (odd) under spatial inversion and odd (even) under transposition in the matrix space of orbitals. 
We note that, in a purely 2D model, the coefficients of all Gell-Mann matrices but $\lambda_{2,4,6}$ vanish by symmetry. 

\begin{table}
\begin{tabular}{|c||c|c|c|c||c|}
\hline 
$i $& $\hat P_x$ & $\hat P_y$&$\hat P_z$ & $\hat R_{\pi/2}$  & examples \\ 
\hline \hline
2 &$ -$ &$-$&$+$&$+$&$s_xs_y(c_x-c_y)$ \\
\hline 
$\Bigg (\begin{array}{c}5\\ 7 \end{array}\Bigg )$ &$ \Bigg (\hspace{-.1cm}\begin{array}{cc}
+ & 0 \\ 
0 & -
\end{array}\hspace{-.1cm}\Bigg ) $&$ \Bigg ( \hspace{-.1cm}\begin{array}{cc}
- & 0 \\ 
0 & +
\end{array}\hspace{-.1cm}\Bigg )  $&$ \Bigg (\hspace{-.1cm}\begin{array}{cc}
- & 0 \\ 
0 & -
\end{array}\hspace{-.1cm}\Bigg )  $&$ \Bigg (\hspace{-.1cm}\begin{array}{cc}
0 & - \\ 
+ & 0
\end{array}\hspace{-.1cm}\Bigg )$&  $\Bigg (\hspace{-.1cm}\begin{array}{c} s_ys_z\\ s_xs_z \end{array}\hspace{-.1cm}\Bigg )$ \\ 
\hline 
\end{tabular} 
\caption{Transformation behavior of coefficients $F^{(i)}_{\omega_n, \v k,-}$, see Eq.~\eqref{eq:ExpFm}, which are odd under $\omega_n \rightarrow - \omega_n$ and under transposition in orbtial space, but even under spatial inversion. The notation is the same as in Table~\ref{tab:ppp}.}
\label{tab:mpm}
\end{table}

\begin{table}
\begin{tabular}{|c||c|c|c|c||c|}
\hline 
$i$ & $\hat P_x$ & $\hat P_y$ &$\hat P_z$ &  $\hat R_{\pi/2}$ & examples \\ 
\hline \hline
0 & $-$ &$ -$&$ -$ &$  -$ &  $s_xs_ys_z$ \\ 
\hline 
1  & $+$ & $+$ &$ -$& $+$&  $s_z, c_{x +y}s_z$ \\ 
\hline
3 &$ -$ &$-$&$ -$&$+$ &$s_xs_ys_z(c_x-c_y)$ \\
\hline
$\Bigg (\begin{array}{c}4\\ 6 \end{array}\Bigg )$ &$ \Bigg (\hspace{-.1cm} \begin{array}{cc}
- & 0 \\ 
0 & +
\end{array}\hspace{-.1cm}\Bigg   ) $&$ \Bigg (\hspace{-.1cm} \begin{array}{cc}
+ & 0 \\ 
0 & -
\end{array}\hspace{-.1cm}\Bigg   ) $&$ \Bigg (\hspace{-.1cm} \begin{array}{cc}
+ & 0 \\ 
0 & +
\end{array}\hspace{-.1cm}\Bigg   ) $&$ \Bigg (\hspace{-.1cm} \begin{array}{cc}
0 & + \\ 
- & 0
\end{array}\hspace{-.1cm}\Bigg   )$& $\Bigg (\hspace{-.1cm} \begin{array}{c} s_x\\ s_y\end{array}\hspace{-.1cm}\Bigg  )$ \\ 
\hline 
8 &$ -$ & $- $&$ -$& $ -$ &  $s_xs_ys_z$  \\ 
\hline 
\end{tabular} 
\caption{Transformation behavior of coefficients $F^{(i)}_{\omega_n, \v k,-}$, see Eq.~\eqref{eq:ExpFm}, which are odd under $\omega_n \rightarrow - \omega_n$ and $\v k \rightarrow -\v k$ but even in orbital space. The notation is the same as in Table~\ref{tab:ppp}.}
\label{tab:mmp}
\end{table}

\subsection{Gell-Mann matrices}
\label{app:GellMann}

The Gell-Mann Matrices are
\begin{equation}
\lambda_0 = \sqrt{\frac{2}{3}} \mathbf 1, \notag
\end{equation}
and
\begin{eqnarray}
\lambda_1 = \left (\begin{array}{ccc}
0 & 1 & 0 \\ 
1 & 0 & 0 \\ 
0 & 0 & 0	
\end{array} \right ), &&
\lambda_2 = \left (\begin{array}{ccc}
0 & -i & 0 \\ 
i & 0 & 0 \\ 
0 & 0 & 0	
\end{array} \right ), \notag  \\
\lambda_3 = \left (\begin{array}{ccc}
1 & 0 & 0 \\ 
0 & -1 & 0 \\ 
0 & 0 & 0	
\end{array} \right ) , &&
\lambda_4 = \left (\begin{array}{ccc}
0 & 0 & 1 \\ 
0 & 0 & 0 \\ 
1 & 0 & 0	
\end{array} \right ), \notag  \\
\lambda_5 = \left (\begin{array}{ccc}
0 & 0 & -i \\ 
0 & 0 & 0 \\ 
i & 0 & 0	
\end{array} \right ), &&
\lambda_6 = \left (\begin{array}{ccc}
0 & 0 & 0 \\ 
0 & 0 & 1 \\ 
0 & 1 & 0	
\end{array} \right ), \notag  \\
\lambda_7 = \left (\begin{array}{ccc}
0 & 0 & 0 \\ 
0 & 0 & -i \\ 
0 & i & 0	
\end{array} \right ) , && 
\lambda_8 =\frac{1}{\sqrt{3}} \left (\begin{array}{ccc}
1 & 0 & 0 \\ 
0 & 1 & 0 \\ 
0 & 0 & -2	
\end{array} \right ).
\end{eqnarray}
These matrices are a complete basis set of the Lie algebra $\mathfrak{u}(3)$ which are orthonormal under the trace form $\tr \lambda_i \lambda_j = 2 \delta_{ij}$.

\section{Generalized t-J model} 
\label{app:tJ}

This appendix includes details on microscopic interactions.

\subsection{Effective interaction in the Cooper channel.}
\label{app:tJIA}

We will consider only the spin singlet channel in which we define
\begin{subequations}
\begin{align}
(\Psi_{\v k}^\dagger)_{mm'} &= \frac{1}{2}[c_{\v k, m, \uparrow}^\dagger c_{-\v k, m', \downarrow}^\dagger - \uparrow \leftrightarrow \downarrow], \\
\quad (\Psi_{\v k})_{m'm} &= \frac{1}{2}[c_{-\v k, m', \downarrow} c_{\v k, m, \uparrow} - \uparrow \leftrightarrow \downarrow].
\end{align}
\end{subequations}
These Cooper pair operators have the property $(\Psi_{\v k}^\dagger)_{mm'} = (\Psi_{-\v k}^\dagger)_{m'm}$. By means of these operators, Eqs.~\eqref{eq:HKanamori} and \eqref{eq:Interactions2} lead to the following interactions in the Cooper channel,
\begin{widetext}
\begin{subequations}
\begin{equation}
H_{\rm IA} \doteq \sum_{mm'} \sum_{\v k, \v k'} V_{\v k, \v k'}^{mm',(S)} (\Psi_{\v k}^\dagger)_{(mm')} (\Psi_{\v k'})_{(mm')}  + \sum_{mm'} \sum_{\v k, \v k'} V_{\v k, \v k'}^{mm',(A)} (\Psi_{\v k}^\dagger)_{[mm']} (\Psi_{\v k'})_{[m'm]} , \label{eq:EffIA}
\end{equation}
where the superscripts $(S)$ ($(A)$) denote symmetry (antisymmetry) under $\v k \rightarrow -\v k$
\begin{align}
V_{\v k, \v k'}^{mm',(S)} &= \bar U+ 2 J_{\rm H,2} (\delta_{ mm'}-1) - \frac{3 J_1^{mm'}}{4} ({(c_{x}+c_{y})(c_{x'}+c_{y'})} + {(c_{x}-c_{y})(c_{x'}-c_{y'}}))  - \frac{3 J_2^{mm'}}{2} ({c_{x}c_y c_{x'}c_{y'} }+ {s_{x}s_y s_{x'}s_{y'}}) \\
V_{\v k, \v k'}^{mm',(A)} &= - \frac{3 J_1^{mm'}}{2}({s_{x}s_{x'}} + {s_{y}s_{y'}}) - \frac{3 J_2^{mm'}}{2}({s_{x}c_y s_{x'}c_{y'}} +{ c_{x}s_y c_{x'}s_{y'}}),
\end{align}
\end{subequations}
\end{widetext}
where $\bar U = U+\frac{3}{2}J_{\rm H,1}$.
Assuming s-wave pairing and exploiting the crystal symmetries in Eq.~\eqref{eq:EffIA} we obtain $V_{\v k, \v k'}^{xz,xz} = V_{\v k, \v k'}^{yz,yz}$ and $V_{\v k, \v k'}^{xz,xy} = V_{\v k, \v k'}^{yz,xy}$. In view of the low power of trigonometric functions, we readily see that only short range pairing is induced. We obtain
\begin{subequations}
\begin{eqnarray}
H_{\rm IA} &=& \sum_{\v k, \v k'} \sum_{i = 0,1,3,5,7,8} V_{\v k, \v k'}^{(i)} \Psi_{\v k}^{(i),\dagger} \Psi_{\v k'}^{(i)} \notag \\
&& +\sum_{\v k, \v k'} V_{\v k, \v k'}^{(0,8)} [\Psi_{\v k}^{(0),\dagger} \Psi_{\v k'}^{(8)} + H.c.] ,
\end{eqnarray}
where
\begin{equation}
\Psi_{\v k}^{(i)} = \sum_{m m'} (\Psi_{\v k})_{mm'} [\lambda_{i}]_{m', m}/2
\end{equation}
and
\begin{eqnarray}
V_{\v k, \v k'}^{(0)} &=& 2 [2V_{\v k, \v k'}^{xz,xz}+V_{\v k, \v k'}^{xy,xy}]/3  \\
V_{\v k, \v k'}^{(8)} &=& 2 [2V_{\v k, \v k'}^{xy,xy}+V_{\v k, \v k'}^{xz,xz}]/3  \\
V_{\v k, \v k'}^{(0,8)} &=& \sqrt{8}[V_{\v k, \v k'}^{xz,xz}-V_{\v k, \v k'}^{xy,xy}]/3  \\
V_{\v k, \v k'}^{(3)} &=& 2 V_{\v k, \v k'}^{xz,xz}  \\
V_{\v k, \v k'}^{(1)} &=& 2 V_{\v k, \v k'}^{xz,yz}  \\
V_{\v k, \v k'}^{(5,7)} &=& 2 V_{\v k, \v k'}^{xz,xy}  .
\end{eqnarray}
\end{subequations}
We note that we only need the s-wave part for $i = 0,8$, d-wave part for $i = 1,3$ and p-wave part for $i = 5,7$. Thus Coulomb repulsion only affects channels $i = 0,8$. This directly leads to the interactions $g_C$ entering the Landau free energy in Eq.~\eqref{eq:LandauIAs}.

\subsection{Kinetic term of the generalized t-J model}
\label{app:tJBands}

\begin{figure}
\includegraphics[scale=.5]{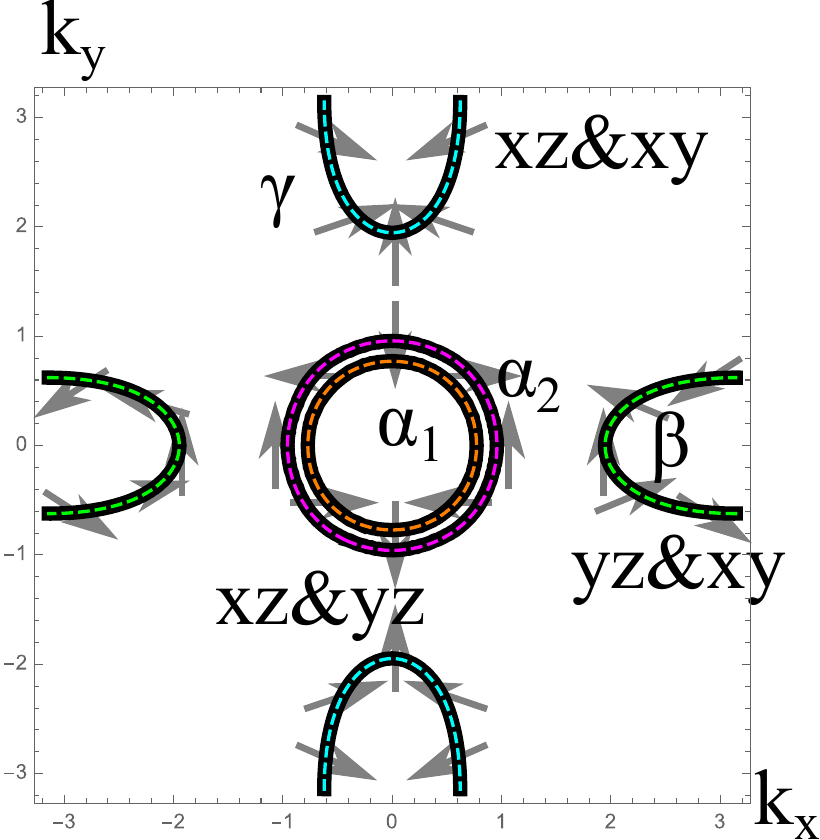} 
\caption{Fermi surfaces for a three band model based on Ref.~\cite{DaghoferDagotto2010}. The black curves correspond to the solution of the full matrix Hamiltonian, colored dashed lines correspond to approximate solutions using the projection on the two orbitals indicated by each of the Fermi surfaces. This picture also includes the vectors $\hat n^{\alpha, \beta, \gamma}$ which rotate in \textit{different} orbital spaces for different Fermi surfaces.}
\label{fig:Fermisurfaces}
\end{figure}

Without repeating details, we replot the Fermi surfaces as suggested by the three orbital tight binding model~\cite{DaghoferDagotto2010} in Fig.~\ref{fig:Fermisurfaces}. Motivated by the excellent numerical agreement due to the large separation of bands near the Fermi surface, we only keep two orbitals near each Fermisurface. The projector on eigenfunctions of the model may then be parameterized by the vectors $\hat n^{\alpha, \beta, \gamma}$ which are also plotted in Fig.~\ref{fig:Fermisurfaces} and account for the rotation of the orbital degrees of freedom around the Fermi surface. Near the two central hole pockets we obtain the projectors on the states in orbital space
\begin{align}
\ket{u_1^{(0)}}\bra{u_1^{(0)}} &= \frac{I_{\alpha,0}+ \hat n_\alpha \mathbf I_\alpha}{2} \; \text{(inner pocket)} ,\\
\ket{u_2^{(0)}}\bra{u_2^{(0)}} &= \frac{I_{\alpha,0}- \hat n_\alpha \mathbf I_\alpha}{2} \; \text{(outer pocket)},
\end{align}
where $ \mathbf I^\alpha = \left (\lambda_1, \lambda_2, \lambda_3\right )$ is the set of Pauli matrices in the space of $xz$ and $yz$ orbitals and $I_{\alpha,0}$ is unity in this space.
At the electron pocket $\beta$, the projector on the band at the Fermi surface is 
\begin{equation}
\ket{u_3^{(0)}}\bra{u_3^{(0)}} = \frac{I_{\beta,0}- \hat n^ \beta \mathbf I_\beta}{2} 
\end{equation}
where $ \mathbf I^\beta = \left (\lambda_6, \lambda_7, \frac{-\lambda_3+\sqrt{3}\lambda_8}{2} \right )$ is the set of Pauli matrices in the space of $yz$ and $xy$ orbitals and $I_{\beta,0}$ is unity in this space. An analogous term may be obtained at the $\gamma$ pocket by a $\pi/2$ rotation, we do not report it here.

We consider a simpler model than Fig.~\ref{fig:Fermisurfaces}, assuming circular Fermi surfaces at $\Gamma,X,Y$ which are parametrized by an angle $\theta$ and $\hat n^\alpha =\left (2c_\theta s_\theta , c^2_\theta- s^2_\theta\right )^T$  in the space $(\ket{xz},\ket{yz})^T$ for the central hole pockets. Furthermore, we assume $\langle \hat n_z^{\beta} \rangle_{FS \beta} = 1/2 , \langle (\hat n_z^{\beta})^2 \rangle_{FS \beta} = 1/3$. This is approximately the same result as one would obtain by a simple ansatz $\hat n^\beta =\left (-2s_\theta , c^2_\theta\right )^T/(1+s_\theta^2)$ in the space $(\ket{yz},\ket{xy})^T$ which is qualitatively in agreement with the orbital structure in Fig.~\ref{fig:Fermisurfaces}. We approximate $c_xc_y \approx \pm 1$ on hole/electron pockets and $s_x s_y \rightarrow k_{F,h}^2 s_\theta c_\theta$ on the hole pockets, $ c_x - c_y \rightarrow k_{F,h}^2 (s_\theta^2-c_\theta^2)/2$ on the hole pockets and $c_x-c_y= -2$ on the electron pocket. We set $k_{F,h} = \sqrt{2/3}$. Then the susceptibility matrix 
\begin{align}
\chi_{\Gamma\Gamma'} &= \frac{1}{2} \sum_{n ,\v k} (\varphi^{\Gamma} (\v k) \lambda_{i(\Gamma)})_{nn}(\varphi^{\Gamma'}(\v k) \lambda_{i(\Gamma')})_{nn} \frac{\tanh\left (\frac{\epsilon_n}{2 T}\right )}{\epsilon_n}  \notag \\
&\simeq \ln(\Lambda/T)[\rho_{h_1} \underline{\tilde \chi}^{(h_1)} + \rho_{h_2} \underline{\tilde \chi}^{(h_2)} +2 \rho_{e} \underline{ \tilde \chi}^{(e)}]_{\Gamma\Gamma'} . \label{eq:susc}
\end{align}
is determined by the following matrices in the space of $\Delta_{\Gamma = 1 \dots 5}$
\begin{align}
\frac{\underline{\tilde \chi}^{(h_1)} + \underline{\tilde \chi}^{(h_2)}}{2} &= \left(
\begin{array}{ccccc}
 \frac{2}{3} & \frac{\sqrt{2}}{3} & \frac{2}{3} & 0 & 0 \\
 \frac{\sqrt{2}}{3} & \frac{1}{3} & \frac{\sqrt{2}}{3} & 0 & 0 \\
 \frac{2}{3} & \frac{\sqrt{2}}{3} & \frac{2}{3} & 0 & 0 \\
 0 & 0 & 0 & \frac{1}{8} & -\frac{1}{24} \\
 0 & 0 & 0 & -\frac{1}{24} & \frac{1}{8} \\
\end{array}
\right) ,\\
\underline{\tilde \chi}^{(e)} &= \left(
\begin{array}{ccccc}
 \frac{2}{3} & -\frac{5}{6 \sqrt{2}} & -\frac{2}{3} & 0 & \frac{1}{\sqrt{6}} \\
 -\frac{5}{6 \sqrt{2}} & \frac{7}{12} & \frac{5}{6 \sqrt{2}} & 0 & -\frac{1}{2 \sqrt{3}} \\
 -\frac{2}{3} & \frac{5}{6 \sqrt{2}} & \frac{2}{3} & 0 & -\frac{1}{\sqrt{6}} \\
 0 & 0 & 0 & 0 & 0 \\
 \frac{1}{\sqrt{6}} & -\frac{1}{2 \sqrt{3}} & -\frac{1}{\sqrt{6}} & 0 & \frac{1}{3} \\
\end{array}
\right). \label{eq:chiHelps}
\end{align}
For simplicity we assume $\rho_{h_1} = \rho_{h_2} \equiv \rho_{h}$ in Fig.~\ref{fig:xDependence} {and we use the kernel of the matrix in the quadratic part of Eq.~\eqref{eq:Landau} to determine $T_c$ at finite $\bar U$}.

\subsection{Renormalization group for the two band model}
\label{app:RG}
{In this appendix, we briefly recapitulate the RG treatment of a two-band model~\cite{Chubukov2015}. Following the discussion of the previous appendix, the Cooper pair creation operator in a given band $n$ takes the form
\begin{equation}
[\Psi_{n}^{(i)}]^\dagger = \frac{1}{2}\sum_{\v k} \lbrace c^\dagger_{\v k, \uparrow, n} c^\dagger_{-\v k, \downarrow, n} - \uparrow \leftrightarrow \downarrow \rbrace (u_{\v k, n}^\dagger \lambda_i u_{\v k, n}).
\end{equation}
Here we used that $u_{-\v k, n}^* = u_{\v k, n}$. For the two band case the situation simplifies as $(u_{\v k, n}^\dagger \lambda_i u_{\v k, n}) \rightarrow 1$. In band space, Cooper interactions take the form
\begin{equation}
H_{\rm int} = \Psi_n^\dagger G_{nn'} \Psi_{n'}.
\end{equation}
As mentioned, for pure Coulomb interaction, $G_{nn'} \propto U$ (independently of $n, n'$). Clearly, the ladder RG describing only particle-particle logarithms (assuming non-degenerate bands) leads to
\begin{equation}
\frac{ dG_{nn'}}{d\ln(D/T)}  = - \sum_{n''} G_{nn''} \rho_{n''} G_{n'' n'}.
\end{equation}
These equations for three separate coupling constants imply the Eqs.~\eqref{eq:RGeqs} along with one conserved coupling constant
\begin{equation}
\frac{d}{d \ln(D/T)} \left (\frac{G_{11} \rho_1 - G_{22} \rho_2}{G_{12} \sqrt{\rho_1 \rho_2}}\right) = 0.
\end{equation}
}


\begin{thebibliography}{57}
\bibitem [{foo()}]{footnote}%
  \BibitemOpen
  \href@noop {} {\bibinfo  {journal} {{In a 5 orbital model there are five
  adjustable parameters, one for each pair angular momentum $L\in [0,4]$.
  However in a simplified 3 orbital model, the pseudo angular momentum of the
  electrons is $l=1$, leading to a total $L\in[0,2]$, giving rise to three
  interaction parameters, allowing the simplified interaction Hamiltonian
  listed here.}}\ }\BibitemShut {NoStop}%
\end{thebibliography}
\end{document}